\documentstyle[prb,aps,eqsecnum,epsf,preprint]{revtex}

\begin{document}

\tightenlines

\title{Spin-wave Scattering in the Effective Lagrangian Perspective}

\author{Christoph P. Hofmann \cite{Adr}}

\address{Department of Physics, University of California at San Diego, 9500
Gilman Drive, La Jolla, California 92093}

\date{May 1998}

\maketitle

\begin{abstract}
\noindent Nonrelativistic systems exhibiting collective magnetic behavior are
analyzed in the framework of effective Lagrangians. The method,
formulating the dynamics in terms of Goldstone bosons, allows to investigate
the consequences of spontaneous symmetry breaking from a unified point of
view. Low energy theorems concerning spin-wave scattering in ferro- and
antiferromagnets are established, emphasizing the simplicity of actual
calculations. The present work includes approximate symmetries and discusses
the modification of the low energy structure imposed by an external magnetic
and an anisotropy field, respectively. Throughout the paper, analogies between
condensed matter physics and Lorentz-invariant theories are pointed out,
demonstrating the universal feature of the effective Lagrangian technique.
\end{abstract}


\section{Introduction}
\label{Intro}

In the following presentation, our interest is devoted to the low energy
analysis of nonrelativistic systems, which exhibit collective magnetic
behavior. We adopt a unified point of view, relying on the method of
effective Lagrangians, and try to understand how the symmetry, inherent in the
underlying theory, manifests itself at low energies. The complex microscopic
description of the systems under consideration is taken into account only
through a phenomenological parametrization, which, in the effective
Lagrangian, emerges in the form of a few coupling constants. Our main concern
will be the question, to what extent in the low energy domain, the actual
structure of quantities of physical interest is dictated by the underlying
symmetry.

Nevertheless, let us first consider the Heisenberg model, which describes the
magnetic systems referred to on a {\it microscopic} level. There, the exchange
Hamiltonian ${\cal H}_0$,
\begin{equation}
\label{Heisenberg}
{\cal H}_0 \ = \ - \, J \, \sum_{n.n.} {\vec S}_m \! \cdot {\vec S}_n ,
\qquad \qquad \qquad J = const. ,
\end{equation}
formulates the dynamics in terms of spin operators ${\vec S}_m$, attached to
lattice sites $m$. Note that the summation only extends over nearest neighbors
and, moreover, the isotropic interaction is assumed to be the same for any two
adjacent lattice sites. According to the sign of the exchange integral $J$,
the above expression leads to an adequate low energy description of systems
exhibiting collective magnetic behavior, both of ferro- and of
antiferromagnets, respectively. In particular, the Heisenberg model is
perfectly suited to study the properties of the excitations near the ground
state -- the spin waves or magnons.

In a more general framework, which represents the starting point of our
systematic approach, these low energy excitations are interpreted as Goldstone
bosons resulting from a spontaneously broken internal symmetry. Indeed, the
Heisenberg Hamiltonian (\ref{Heisenberg}) is invariant under a simultaneous
rotation of the spin variables, described by the symmetry group G = O(3),
whereas the ground state of a ferromagnet, e.g., breaks this symmetry
spontaneously down to H = O(2): all the spins are aligned in one specific
direction, giving rise to a nonzero spontaneous magnetization. Although the
antiferromagnetic ground state does not display spontaneous magnetization, it
also spontaneously breaks the symmetry. Unlike for a ferromagnet, its
microscopic description is highly nontrivial -- in our analysis, we assume the
same internal symmetry breaking pattern to inhere in this system as well:
$\mbox{G = O(3)} \to \mbox{H = O(2)}$.

Whenever a physical system exhibits spontaneous symmetry breaking and,
furthermore, the corresponding Goldstone bosons represent the only low energy
excitations without energy gap, we do have a very powerful means at our
disposal to analyze its low energy structure: chiral perturbation theory
($\chi$PT). The method was originally developed in connection with
Lorentz-invariant field theories \cite{Weinberg junior,Coleman Callan,Li
Pagels,Weinberg seminal,Gasser Leutwyler QCD}, admitting, in particular, a low
energy analysis of the strong interaction, described by quantum chromodynamics
(QCD). $\chi$PT has also proven to be very useful in the investigation of other
systems where Goldstone bosons occur (see e.g. \cite{Hasenfratz
Leutwyler,Hasenfratz Niedermayer 1991,Hasenfratz Niedermayer 1993}).

In condensed matter physics, spontaneous symmetry breaking is a common
phenomenon and effective field theory methods are widely used in this domain.
Only recently, however, has chiral perturbation theory been extended to such
nonrelativistic systems \cite{Leutwyler NRD,Leutwyler Phonons,Soto,Burgess},
demonstrating its applicability to solid state physics as well; especially the
ferro- and the antiferromagnet, the systems to be examined below, may be
analyzed in the framework of $\chi$PT. The method is based on effective
Lagrangians which exploit the symmetry properties of the underlying theory,
i.e. the Heisenberg model in our case, and permits a systematic low energy
expansion of quantities of physical interest in powers of inverse wavelength.

While in an anomaly free, Lorentz-invariant field theory an invariance theorem
\cite{Leutwyler foundations} guarantees that the effective Lagrangian will
inherit the symmetries of the underlying model, the same statement is
no longer true in the nonrelativistic domain: terms of topological nature
happen to occur in the effective description, the corresponding Lagrangian
being G-invariant only up to a total derivative
\cite{Leutwyler NRD,Soto,Fradkin}.
Especially for ferromagnets, a term connected with the Brouwer degree emerges
which is not invariant under the group G = O(3), whereas an analogous
contribution is absent in the effective Lagrangian of an antiferromagnet --
as we will see, the main differences in the low energy behavior of these two
systems are a consequence of this striking fact.

The whole analysis concerns the properties of magnetic systems at wavelengths
large compared to the intrinsic scales of the theory, i.e. to the lattice
spacing $a$ -- the effective theory does not resolve the lattice structure of
the system, i.e. refers to the continuum limit. Clearly then, the effective
Lagrangian method does not admit to discuss the physics of a solid body on a
microscopic scale. Rather, we are interested in how the actual structure of
several low energy phenomena encountered in magnetic systems can be
interpreted as an immediate consequence of the hidden symmetry. The method has
proven to be very efficient in other areas, above all in analyzing the low
energy behavior of the strong interaction; in particular, the effective QCD
Lagrangian allows to perform a concise derivation of certain low energy
theorems concerning the pions, which represent the Goldstone bosons in this
relativistic sector. The main intention of the following work is to
demonstrate, that chiral perturbation theory, extended to nonrelativistic
systems, is an equally powerful tool.

One principal result of the present paper will be the establishment of
low energy theorems concerning the scattering amplitude of ferro- and
antiferromagnetic spin waves, respectively. The straightforward effective
calculation, as opposed to the complicated microscopic analysis, exhibits the
efficiency of the method, which, moreover, can systematically be extended to
higher orders of momentum \cite{Leutwyler Phonons,Soto}.
Likewise, it is not a complicated matter to include
a weak external magnetic or an anisotropy field, respectively, into the
effective machinery, in order to discuss the modifications thereby imposed on
the low energy structure.

As far as the ferromagnet is concerned, our continuum approach makes contact
with an important result to be found in the literature: Dyson, in his thorough
microscopic analysis of a cubic ferromagnet within the Heisenberg model,
calculated the scattering cross section regarding ferromagnetic spin waves
more than four decades ago \cite{Dyson1}. The fact that our result coincides
with his, may be viewed as some kind of a test run for the applicability of
chiral perturbation theory in the present context: the interaction among
ferromagnetic spin waves is described correctly in this new framework.

In a sense to be specified below, the leading order effective Lagrangian of an
antiferromagnet closely resembles the one describing QCD at lowest order. As a
consequence, many results concerning chromodynamics can be adopted to
antiferromagnets, the corresponding low energy phenomena manifesting
themselves in analogous ways. This feature of universality offers the
opportunity to discuss certain phenomena well-known in Lorentz-invariant
theories in the different language of solid state physics and vice versa.
Indeed, throughout the paper we will make quite often use of such comparisons
and analogies, in order to make the material easily accessible, both to solid
state physicists and to the relativistic community.

For the sake of selfconsistency of the present work, we give a brief outline
of the main ideas of chiral perturbation theory and review the effective
description of ferro- and antiferromagnets, leaning thereby on references
\cite{Leutwyler NRD,Leutwyler foundations,Leutwyler Brasil}. In contrast to
the analysis found therein, our approach tries to adopt "magnetic language",
paving the way to confront our theoretical findings with the microscopic
analysis.

\section{The Nonrelativistic Domain}
\label{NRD}

In the following two sections, we analyze systems exhibiting collective
magnetic behavior with respect to their symmetry properties, referring both to
space-time and to internal transformations. Special emphasis is put into the
internal symmetry G = O(3) inherent in the Heisenberg
model, which is spontaneously broken by the ground state of the corresponding
magnetic systems. Discussing the consequences resulting from this breaking in
the general framework of Goldstone's theorem, we try to work out the main
differences of the low energy structure between Lorentz-invariant theories and
the nonrelativistic domain.

As far as {\it space-time} symmetries are concerned, we are faced with the
following situation in a continuum description of condensed matter: the object
under investigation, e.g. a magnetic crystal, singles out a preferred frame of
reference, the rest frame. In contrast to relativistic theories, where the
vacuum is invariant under Lorentz transformations or, more generally, under
the whole Poincar\'e group, the ground state of a solid fails to be invariant.
As a consequence, the statement that the vacuum expectation value of a vector
operator $A^{\mu}$ has to vanish, no longer holds: the time component $A^0$
may pick up an expectation value in the ground state. This observation
represents an essential ingredient of the whole low energy analysis in the
nonrelativistic domain, since such nonzero quantities can acquire the role of
order parameters. As we soon will see, the ferromagnet is such a
nonrelativistic system, where $A^0$ is the time component of a conserved
current.

As the effective analysis refers to large wavelengths, it does not resolve the
microscopic structure of a solid and the system hence appears homogeneous.
Accordingly, the effective Lagrangian is invariant with respect to
translations. On the other hand, the effective Lagrangian is not invariant
under rotations, since the lattice structure of a solid singles out preferred
directions. In the case of a cubic lattice, the anisotropy, however, only
shows up at higher orders of the derivative expansion \cite{Hasenfratz
Niedermayer 1993}.\cite{footnote1} In the following discussion, we assume that
our magnetic systems exhibit this type of lattice structure: the underlying
theory is the Heisenberg model of a cubic ferro- and antiferromagnet,
respectively. Under this assumption, the leading order effective Lagrangians
relating to are then invariant both under translations and under rotations.

Let us now turn to {\it internal} symmetries. In addition to the group
R = O(3), which refers to rotations in three-dimensional Euclidean space, a
further symmetry group O(3) comes into play, associated with the isotropic
exchange interaction in the Heisenberg model: G = O(3). Note that this group
corresponds to internal symmetry transformations in the space of the spin
variables.

Invariance of the Heisenberg Hamiltonian ${\cal H}_0$ (\ref{Heisenberg}) with
respect to the Lie group G = O(3), characterized by the generators
$Q_i$,
\begin{equation}
[Q_i, {\cal H}_0] = 0 ,
\end{equation}
gives rise to three conserved currents $J^{\mu}_i(x)$,\cite{footnote2}
\begin{equation}
\partial_{\mu} J^{\mu}_i(x) \, \equiv \, \partial_0 J^0_i(x) +
\partial_r J^r_i(x) \, = \, 0 .
\end{equation}
The generators $Q_i$ are space integrals over the corresponding charge
densities $J^0_i(x)$,
\begin{equation}
Q_i = \int \!\! d^3x J^0_i(x) ,
\end{equation}
obeying the commutation relations
\begin{equation}
[Q_i, Q_j] = i \varepsilon_{ijk} \, Q_k .
\end{equation}
The exact symmetry G = O(3) is spontaneously broken down to H = O(2): whereas
the Hamiltonian of the theory is invariant under the full group G, the ground
state of the system is invariant only under the subgroup H.

In a microscopic description of a ferromagnet, e.g., this statement shows up
as follows. The generators of the symmetry group are given by the sum over all
spins,
\begin{equation}
Q_i = \sum_n S^i_n .
\end{equation}
The commutation rule
\begin{equation}
[S^i_m, S^j_n] \, = \, i \delta_{mn} \, \varepsilon_{ijk} \, S^k_m
\end{equation}
insures that the Heisenberg Hamiltonian is invariant under G. The ground state
of a ferromagnet, on the other hand, in which all the spins are aligned in one
specific direction, let's say along the positive $3^{rd}$ axis in spin space,
is invariant only under H = O(2), represented by the single
generator $Q_3 = {{\sum}_{n}}S^3_n$.

At this point, the microscopic analysis makes contact with the continuum
approach: the third component of the operator of the total spin,
${{\sum}_{n}}S^3_n$, is related to the third component of the charge density
operator, $J^0_3$, by
\begin{equation}
{\sum_{n}} S^3_n = \int \!\! d^3x J^0_3(x) .
\end{equation}
Taking the vacuum expectation value on either side of this equation, we arrive
at
\begin{equation}
NS \; = \; \langle 0|\, J^0_3 \,|0 \rangle \, V ,
\end{equation}
where $N$ denotes the total number of lattice sites, $S$ is the highest
eigenvalue of the spin operator $S^3_n$, and $V$ is the volume of the entire
crystal. Accordingly, the vacuum expectation value of the third component of
the charge density operator is nonzero,
\begin{equation}
\langle 0|\, J^0_i \,|0 \rangle \ = \ \delta^3_i \; \frac{NS}{V} \ = \
\delta^3_i \, \Sigma ,
\end{equation}
to be identified with the spontaneous magnetization $\Sigma$. This
quantity represents the most prominent order parameter in the description of
a ferromagnet, its nonzero value signaling spontaneous symmetry breaking
-- let us elaborate this statement a bit further.

If the ground state of a ferromagnet was symmetric with respect to the whole
group G = O(3), none of the operators $J^0_i(x)$, which transform in a
nontrivial manner under G, could develop a vacuum expectation value different
from zero. The fact that $J^0_3(x)$ nevertheless does so, indicates, that the
ground state of a ferromagnet must single out some specific direction in the
internal space of the spin variables: being symmetric only with respect to the
subgroup H = O(2), the ground state does not share the full symmetry of the
Hamiltonian -- therefore, $\langle 0|\, J^0_3 \,|0 \rangle \neq 0$ may be viewed
as a quantitative measure of spontaneous symmetry breaking. Generally, nonzero
vacuum expectation values of local operators, which transform in a nontrivial
manner under a symmetry group G, are referred to as order parameters.

The above analysis exhibits, that the ferromagnet represents a physical
system, where the most prominent order parameter is associated with the time
component of a conserved current. Comparing this situation with the one in an
antiferromagnet, we realize that, in this case, the spontaneous magnetization
happens to vanish, $\langle 0|\, J^0_i \,|0 \rangle = 0$. Although the
symmetries of the ground state would have nothing against
$\langle 0|\, J^0_3 \,|0 \rangle$ taking on the role of an order parameter,
this possibility is ruled out for dynamical reasons. As far as the
antiferromagnet is concerned, the so-called staggered magnetization,
$\Sigma_s$, turns out to be the most important order parameter -- this
quantity, however, is not associated with the time component of a conserved
current.

\section{Goldstone Theorem}
\label{GT}

An essential feature of the present low energy analysis is the occurrence of
spontaneous symmetry breaking. The consequences of this phenomenon for the
level spectrum of the corresponding systems are dictated by Goldstone's
theorem.

Let us first consider its {\it relativistic} version
\cite{Goldstone,Coleman,Guralnik Hagen Kibble} for arbitrary Lie groups G and
H, associated with an internal symmetry. In the absence of gauge fields,
spontaneous symmetry breaking in a Lorentz-invariant theory implies the
existence of massless particles, whose number, $n_{GB}$, is determined by
the dimension of the coset space G/H: $n_{GB}$ = dim(G) -- dim(H).
The current operators $J^{\mu}_a$ referring to G/H, $a = 1 \ldots n_{GB}$,
couple to the vacuum, the corresponding vacuum-to-Goldstone boson matrix
elements being nonzero. In QCD, e.g., the axial current, $J^{5 \mu}_a$,
displays this property,\cite{footnote3}
\begin{equation}
\langle 0|\, J^{5 \mu}_a \, |{\pi}^{b}(\vec k) \rangle = i {\delta}^b_a k^{\mu}
F .
\end{equation}
The {\it nonrelativistic} version of the theorem \cite{Guralnik Hagen
Kibble,Lange,Chadha Nielsen} is weaker. In the absence of long-range forces,
spontaneous symmetry breaking in a nonrelativistic system leads to low energy
excitations, whose frequency $\omega$ tends to zero for $\vec{k} \rightarrow
0$. In contrast to the relativistic version, the theorem does now neither
specify the exact form of the dispersion relation at large wavelengths, nor
does it determine the number of different Goldstone {\it particles}: these
features of the Goldstone degrees of freedom are not fixed by symmetry
considerations alone -- rather, in the case of a Lorentz-noninvariant ground
state, they depend on the specific properties of the corresponding
nonrelativistic systems. Only the number of real Goldstone {\it fields} turns
out to be universal, given again by the dimension of G/H.

As far as the matrix elements of the operators $J^{\mu}_i$ between the vacuum
and the Goldstone states are concerned, we find the following situation in the
nonrelativistic domain: two independent coefficients have to be introduced, in
order to characterize the matrix elements of the charge densities and the
currents in question. Furthermore, the number of independent Goldstone states,
labeled by the index $n$, $|{\pi}^n \rangle$, remains open. Quite generally,
we may write
\begin{equation}
\label{GTmatrix}
\langle 0|\, J^0_i(x) \, |{\pi}^n \rangle = i C^n_i(\vec k) \, e^{-ikx} ,
\qquad \langle 0|\, J^r_i(x) \, |{\pi}^n \rangle = i D^{n r}_i \! (\vec k) \,
e^{-ikx} .
\end{equation}
The two quantities are related via current conservation, leading to the
dispersion law\cite{footnote4}
\begin{equation}
\label{GTdisp}
C^n_i(\vec k) \; \omega \; = \; D^{n r}_i \! (\vec k) \, {\vec k} .
\end{equation}
Note that the exact form of the dispersion relation has not yet been
specified: symmetry alone does not allow to determine the explicit ${\vec
k}$-dependence of the coefficients $C^n_i(\vec k)$ and $D^{n r}_i \! (\vec k)$
-- rather, their actual structure depends on the specific properties of the
nonrelativistic systems under consideration. This is to be compared with the
Lorentz-invariant situation, where the ratio of the energy to the momentum is
universal, determined by the velocity of light, ${\omega}^2 = c^2 {\vec k}^2$:
every Goldstone boson turns out to be massless, provided that G is an exact
symmetry of the Lagrangian.

From a theoretical analysis of magnetic systems, e.g. based on the Heisenberg
model, as well as from the experimental side, e.g. from neutron scattering, it
is well-known that the structure of the ferromagnetic dispersion relation is
quite different from the antiferromagnetic one: at large wavelengths, the
former takes a {\it quadratic} form, whereas the latter follows a {\it linear}
law. The mechanism which leads to this pattern and, at the same time, explains
the different number of independent magnon states -- {\it one} for a
ferromagnet, {\it two} for an antiferromagnet -- is understood \cite{Kittel
QT,Landau Lifshitz,Randjbar-Daemi Salam Strathdee}. Remarkably, in the
framework of our effective description, the difference in the value of a
single observable, the spontaneous magnetization, suffices to answer both
questions: the one concerning the number of independent magnon states as well
as the one referring to their dispersion law \cite{Leutwyler NRD}. We shall
briefly review the chain of arguments in a later paragraph, once we have the
corresponding effective Lagrangians at our disposal.

In the following analysis, the microscopic structure of the system does not
play a significant role. A brief discussion of the spin-wave excitations
within the Heisenberg model of a ferromagnet may be found in appendix A, which
also tries to give a intuitive understanding of Goldstone's theorem in the
present context.

\section{Aspects of Chiral Perturbation Theory}
\label{XPT}

Chiral perturbation theory is an efficient method to analyze the low energy
structure of systems with a spontaneously broken symmetry. It is an effective
theory, formulated in terms of Goldstone fields, applicable both to
Lorentz-invariant theories and to the nonrelativistic domain. An
essential condition for the whole framework to be consistent is the validity
of three assumptions, whose significance in connection with magnetic systems,
we are now going to examine in succession.

The {\it first} one supposes, that the magnons are the only excitations
without energy gap. A more realistic description of a magnet faces the fact
that the system admits other excitations with this property. In particular,
phonons occur, representing the Goldstone bosons generated by the spontaneous
breaking of translation invariance. We will concentrate on the magnons and
disregard all other degrees of freedom -- the same idealization has been used
by Dyson and many others.

The exchange of magnons leads to singularities in the low energy region,
particularly to poles occurring in the time-ordered correlation functions of
the currents and charge densities. In the two-point function $\langle 0|\,
T\{J^r_i(x)J^s_k(0)\} \,|0 \rangle$, e.g., the pole term arises from the
exchange of a magnon between the two currents: the first current emits a magnon
which propagates and gets absorbed by the second one. Apart from this
one-magnon exchange, multimagnon exchange processes, corresponding to branch
points, also occur and complicate the analysis considerably. At this moment,
however, a {\it second} assumption, known as the pion pole dominance hypothesis,
comes into play: one postulates that the singularities due to one-magnon
exchange dominate the low energy expansion.

A {\it third} ingredient of the low energy analysis is the assumption that the
residues of the pole terms, i.e. the vertices representing the interaction
among the magnons, do admit an expansion in powers of inverse wavelength. Note
that the correlation functions themselves, of course, cannot be expanded in
this way, due to the pole terms showing up therein. This third assumption is
essential in chiral perturbation theory: it allows to analyze the low energy
structure of scattering amplitudes, form factors and other quantities of
physical interest in a systematic manner.

In view of some applications to be presented later on, it is convenient to
make use of the external field technique: one considers the response of the
system to perturbations generated by suitable external fields $f^i_{\mu}(x)$,
coupled to the currents $J^{\mu}_i(x)$. All the various correlation functions
are collected compactly in a generating functional $\Gamma \{ f \}$,
\begin{equation}
\label{XPTgenFunct}
e^{i \Gamma \{ f \}} = \sum_{n=0}^{\infty} \frac{i^n}{n!} \int \! \! d^4x_1...
d^4x_n \, f^{i_1}_{{\mu}_1}(x_1)... f^{i_n}_{{\mu}_n}(x_n) \langle 0|\,
T \{ J^{{\mu}_1}_{i_1}(x_1)... J^{{\mu}_n}_{i_n}(x_n) \} \,|0 \rangle ,
\end{equation}
where the external fields, $f^i_{\mu}(x)$, merely serve as auxiliary variables:
appropriate functional derivatives of $e^{i\Gamma\{f\}}$ with respect to these
quantities reproduce the correlation functions referred to above. The
generating functional describes the transitions which occur when the system is
perturbed by an external field, ${\cal H}\rightarrow{\cal H} - \int\!\! d^3x
f^i_{\mu} J^{\mu}_i$, where ${\cal H}$ represents the Hamiltonian of the
theory. In particular, $e^{i\Gamma\{f\}}$ is the probability amplitude for the
system to remain in the ground state for $t \rightarrow + \infty$, if it was
there at $t \rightarrow - \infty$.

Up to this point, the discussion of magnetic systems was based on a
spontaneously broken internal symmetry, G = O(3), inherent in the {\it
underlying} theory, the Heisenberg model. The basic idea now in constructing
an {\it effective} theory is to interpret the one-particle reducible graphs
occurring in the underlying theory as tree graphs of an effective field
theory, which involves magnon fields as fundamental variables: magnons are to
be described by two scalar fields, denoted by ${\pi}^a(x), a = 1,2$, and the
pole terms generated by one-magnon exchange, e.g., arise now from magnon-field
propagators.

In this language, the expansion of the vertices in powers of inverse wavelength
corresponds to a derivative expansion of the effective Lagrangian. The
translation of the various vertices into the corresponding terms of the
effective Lagrangian is trivial: the one describing the interaction between
four magnons, e.g., is represented through a term containing four magnon
fields, together with a not yet specified number of space- and
time-derivatives. In addition to the purely magnonic vertices, describing the
interaction of magnons among themselves, the effective Lagrangian also
contains contributions involving the external fields, which describe the
transitions generated by the perturbation $f^i_{\mu}J^{\mu}_i$. The matrix
element $\langle 0|\, f^i_{\mu}J^{\mu}_i|{\pi}^n \rangle$, e.g., which
representsthe probability amplitude for the external field to excite one of the
magnon states, is represented through a term linear in the fields
$f^i_{\mu}(x)$, ${\pi}^a(x)$. The effective Lagrangian thus merely collects the
information about the various vertices occurring in the underlying theory.

In switching from the underlying theory over to the effective Lagrangian, one
could think, at first sight, that the latter would simply inherit the former's
internal symmetry: in connection with magnetic systems, one would therefore be
inclined to construct an effective Lagrangian out of O(3)-invariant
expressions of increasing complexity -- respecting, of course, the symmetry
properties under space-time transformations, i.e. invariance under
translations and space rotations in the present context.

This plausible way of proceeding, however, does not generally lead to correct
effective Lagrangians \cite{Leutwyler NRD,Soto,D'Hoker Weinberg}. A detailed
analysis of the low energy structure of nonrelativistic systems shows
\cite{Leutwyler NRD}, that the leading order effective Lagrangian of a
ferromagnet indeed is not invariant under the group G = O(3) of the Heisenberg
model. Note that this peculiarity is specific to the nonrelativistic domain
and does not show up in an anomaly free, Lorentz-invariant theory -- there,
the effective Lagrangian can always be brought to a G-invariant form
\cite{Leutwyler foundations}.

Once the explicit effective Lagrangian at hand, chiral perturbation theory
exhibits its full strength, emphasizing the simplicity of actual calculations.
Moreover, the method allows to systematically take into account interactions,
which {\it explicitly} break the symmetry of the underlying theory, provided
that they can be treated as perturbations. As far as our magnetic systems are
concerned, we will investigate the effect of an external magnetic and an
anisotropy field on the structure of various low energy phenomena.

\section{Low Energy Behavior of Ferro- and Antiferromagnets}
\label{F/AF}

We now confine our attention to the low energy properties of ferro- and
antiferromagnets. First of all, the corresponding effective Lagrangians have
to be written down. According to Goldstone's theorem, spontaneous symmetry
breaking of the rotation group inherent in the Heisenberg model,
$\mbox{G = O(3)} \to \mbox{H = O(2)}$, gives rise to two real magnon fields.
It is convenient to use a covariant representation for the magnon field,
replacing the two variables ${\pi}^1$, ${\pi}^2$ by a three-dimensional unit
vector $\vec{U} = (U^1,U^2,U^3)$, which transforms with the vector
representation of G = O(3).

In this notation, the leading order effective Lagrangian of a {\it
ferromagnet} reads \cite{Leutwyler NRD}
\begin{equation}
\label{LeffFerro}
{\cal L}_{eff}^{F} \ = \ \Sigma \, \frac{{\partial}_0 U^1U^2 - {\partial}_0
U^2U^1}{1 + U^3} \ + \ {\Sigma}f^i_0U^i \ - \ \mbox{$ \frac{1}{2}$} F^2
D_rU^iD_rU^i ,
\end{equation}
the last term being proportional to the square of the covariant derivative of
$\vec U$,
\begin{equation}
D_rU^i \, = \, {\partial}_rU^i + \varepsilon_{ijk} f^j_rU^k .
\end{equation}
At leading order of the low energy expansion, the ferromagnet is thus
characterized by two different low energy coupling constants, $\Sigma$ and
$F$. The first term is related to a topological invariant, to the Brouwer
degree of the map ${\vec U}(\pi)$. Remarkably, due to this contribution, which
does not involve the auxiliary field $f^i_0$, the effective Lagrangian of a
ferromagnet fails to be invariant under the group G = O(3). The second term in
(\ref{LeffFerro}) exhibits the same coupling constant, the spontaneous
magnetization. Note that these two expressions, proportional to the order 
parameter, would not be permitted in Lorentz-invariant effective theories --
they represent the main novelty occurring in condensed matter physics, where
nonrelativistic kinematics is less restrictive than Lorentz invariance.

The ground state of an antiferromagnet, on the other hand, does not exhibit
spontaneous magnetization, such that the above two contributions do not show up
in the effective description of this system. The explicit expression for the
leading order effective Lagrangian of an {\it antiferromagnet} is given by
\begin{equation}
\label{LeffAF}
{\cal L}_{eff}^{AF} \, = \, \mbox{$ \frac{1}{2}$} F^2_1
D_0U^iD_0U^i - \mbox{$ \frac{1}{2}$} F^2_2 D_rU^iD_rU^i , \quad
D_{\mu}U^i = {\partial}_{\mu}U^i + \varepsilon_{ijk} f^j_{\mu}U^k .
\end{equation}
As it is the case in the relativistic domain, the effective Lagrangian is
invariant with respect to the symmetry group G. Since the expression
(\ref{LeffAF}) gives rise to a linear dispersion relation, it is more
convenient to count energies as quantities of the same order as momenta,
$\omega \propto |{\vec k}|$, rather than organizing the bookkeeping according
to $\omega \propto {\vec k}^2$, as for the ferromagnet before. The Lagrangian
also contains two effective coupling constants, $F_1$ and $F_2$. Note
that the contribution involving $F^2_2$ represents the analog of the
$F^2$-term in (\ref{LeffFerro}), whereas the first contribution, proportional
to $F^2_1$, would appear in the effective Lagrangian of a ferromagnet only at
subleading order.

Having the explicit leading order effective Lagrangians for ferro- and
antiferromagnets at hand, we are now capable of describing the low energy
behavior of these two systems.\cite{footnote5}

Let us start with the ferromagnet, whose ground state displays a nonzero
spontaneous magnetization. This specific information on the system suffices to
determine the corresponding leading order effective Lagrangian within our
framework, which is characterized by the groups $\mbox{G = O(3)} \to \mbox{H =
O(2)}$ (spontaneous symmetry breaking) and R = O(3) (invariance under space
rotations), respectively. The associated equation of motion is the
Landau-Lifshitz equation, well-known in solid state physics, which describes
the dynamics of ferromagnetic spin waves. Its nonrelativistic,
Schr\"odinger-type structure -- first order in time, but second order in space
-- for its part determines the number of independent magnon states: as only
positive frequencies occur in its Fourier decomposition, a complex field is
required to describe one particle -- in a ferromagnet there exists only {\it
one} type of spin-wave excitation. Remember that, in the nonrelativistic
domain, Goldstone's theorem is too weak to make such a statement: it only
predicts the number of real magnon {\it fields}, dim\{O(3)/O(2)\} = 2, but
leaves open the number of different magnon {\it particles}. Moreover, merely
claiming that the frequencies $\omega$ must tend to zero for ${\vec k} \to 0$,
the theorem does not quantitatively specify the dispersion relation at large
wavelengths. Its quadratic form, resulting from the effective Lagrangian
(\ref{LeffFerro}),
\begin{equation}
\label{disprelF}
\omega(\vec k) \, = \, \gamma{\vec k}^2 + {\cal O}(|{\vec k}|^4) , \qquad
\quad \gamma \equiv \frac{F^2}{\Sigma} ,
\end{equation}
is a consequence of the Euclidean symmetry R = O(3) as well as of the specific
information on the ground state of the ferromagnet, concerning the nonzero
value of its spontaneous magnetization.

The effective machinery, relying on the external field technique, may now be
put in operation, providing us with a derivative expansion of the correlation
functions needed: the Landau-Lifshitz equation is to be solved iteratively and
the respective solutions for the magnon field $U^a$ are to be inserted into the
effective Lagrangian (\ref{LeffFerro}). At leading order, the whole
information on the correlation functions is then collected compactly in the
generating functional, $\Gamma \{f\} {|_{tree}} = \int \!\!d^4x{\cal
L}_{eff}^{F}$, and may be obtained by taking appropriate derivatives with
respect to the auxiliary fields $f$. Our interest is now devoted to the
contribution proportional to $f^a_0f^b_0$, i.e. to the two-point function of
the charge densities, $\langle 0|\, T \{ J^0_a J^0_b \} \,|0 \rangle$, for it
is this quantity which allows us to calculate the matrix element
$\langle 0|\, J^0_a |\pi(\vec k)\rangle$. The result is
\begin{equation}
\label{FerroMatrixTime}
\langle 0|\, J^0_a |\pi(\vec k) \rangle = {\varepsilon}_a \sqrt{\Sigma} ,
\qquad {\varepsilon}_a = \mbox{$ \frac{1}{\sqrt{2}}$} (1,-i) .
\end{equation}
Current conservation and invariance under R = O(3) determine the
corresponding spatial expression,
\begin{equation}
\label{FerroMatrixSpace}
\langle 0|\, J^r_a |\pi(\vec k) \rangle \, = \, {\varepsilon}_a k^r \gamma
\sqrt{\Sigma} \, = \, {\varepsilon}_a k^r \, F^2 / \sqrt{\Sigma} .
\end{equation}
At this stage of the effective analysis, we may look back to the general
expressions (\ref{GTmatrix}) for these matrix elements. Within the effective
framework, the explicit structure of the two quantities $C^n_i(\vec k)$ and
$D^{n r}_i \! (\vec k)$ has now been determined. For ferromagnets, there exists
only one polarization state, $|\pi(\vec k) \rangle \Leftrightarrow |{\pi}^n
\rangle, \; n = 1$. In particular, the coefficient $C^n_a(\vec k)$ does not
dependent on momentum -- equation (\ref{GTdisp}) then takes the quadratic form
(\ref{disprelF}).

As far as the antiferromagnet is concerned, quite a different low energy
description emerges, because, for this system, the spontaneous magnetization
happens to vanish. The corresponding equation of motion is of second order
both in space and in time, its relativistic structure determining the number
of independent magnon states: the Fourier decomposition contains both positive
and negative frequencies, such that a single real field suffices to describe
one particle. Accordingly, there exist {\it two} different types of spin-wave
excitations in an antiferromagnet -- as it is the case in Lorentz-invariant
theories, Goldstone fields and Goldstone particles are in one-to-one
correspondence. Moreover, these excitations now follow a linear dispersion
relation,
\begin{equation}
\label{disprelAF}
\omega(\vec k) \, = \, v|{\vec k}| + {\cal O}(|{\vec k}|^2) , \qquad \quad v
\equiv \frac{F_2}{F_1} ,
\end{equation}
corresponding to a massless particle moving with velocity $v$.

The transition matrix elements of the charge densities and currents take the
specific form
\begin{equation}
\label{AFmatrix}
\langle 0|\, J^0_a |{\pi}^b(\vec k) \rangle = i {\delta}^b_a |\vec{k}| F_2 /
{\sqrt{2\omega}} , \qquad \langle 0|\, J^r_a |{\pi}^b(\vec k) \rangle =
i {\delta}^b_a k^r v F_2 / {\sqrt{2\omega}} .
\end{equation}
There are now two polarization states, $|{\pi}^a(\vec k) \rangle, \; a = 1,2
\Leftrightarrow |{\pi}^n \rangle, \; n = 1,2$, which are associated with the
operators $J^0_a, J^r_a$, referring to the coset space G/H. Unlike for
ferromagnets, the coefficient $C^n_a(\vec k)$ does depend on momentum:
equation (\ref{GTdisp}) then leads to the linear dispersion law
(\ref{disprelAF}).

We would like to emphasize, once more, that this striking difference in the low
energy behavior of ferro- and antiferromagnets, cannot be understood in terms
of symmetry considerations: with respect to internal as well as space-time
symmetries, the two systems are identical in our effective framework. Rather,
the difference originates from the actual value of a {\it single} effective
coupling constant, the spontaneous magnetization. The number of independent
magnon states, the form of their dispersion relation, the low energy
representation of scattering amplitudes, $\ldots$ -- the explicit appearance
of all these low energy phenomena can be traced back to the different behavior
of the respective ground states.

\section{Effective Coupling Constants}
\label{Couplings}

In the nonrelativistic domain the manifold of effective coupling constants is
larger than in Lorentz-invariant theories. In connection with our systems
exhibiting collective magnetic behavior, the situation is the following: both
for ferro- and for antiferromagnets there are two couplings to be determined at
leading order: $\Sigma$ and $F$ for a ferromagnet, $F_1$ and $F_2$
for an antiferromagnet. Part of this information, as we will see, may be
obtained from the dispersion relation.

In relativistic theories, Lorentz symmetry imposes a universal law,
${\omega}^2 = c^2 {\vec k}^2$, independent of the specific properties of the
system under consideration. In the nonrelativistic domain, effective constants
happen to show up in the leading order dispersion relation: for 
ferromagnets we obtained $\omega = \gamma{\vec k}^2 \equiv (F^2/\Sigma){\vec
k}^2$, for antiferromagnets we got $\omega = v|{\vec k}| \equiv
(F_2/F_1)|{\vec k}|$. So, on the one hand, the nonrelativistic situation is
more complex: less information on the systems is available via symmetry, such
that a larger number of constants has to be fixed phenomenologically. On the
other hand, those combinations of low energy constants which happen to appear
in the dispersion law, are comparatively easy to determine by experiment: the
respective coefficients, $\gamma$ and $v$, may be obtained by scattering
neutrons on a given magnetic crystal.

As far as the ferromagnet is concerned, the spontaneous magnetization is
easily accessible as well -- the two low energy couplings, $\Sigma$ and
$F$, occurring in the leading order effective Lagrangian ${\cal
L}^{F}_{eff}$, are thus fixed. For the antiferromagnet, which does not
develop such a vacuum expectation value, the situation is more delicate. So
far, we have merely been determining the ratio $v = {F_2}/{F_1}$.

Here comes the appropriate place, where we may pause for a moment and
deviate into the field of the strong interaction. The point is that a close
resemblance between the leading order effective Lagrangian of an
antiferromagnet and the one describing QCD is observed. Since we know that the
low energy sector of the strong interaction is successfully described within
chiral perturbation theory, it might prove to be instructive to have a look at
the specific way the low energy couplings are determined there. Maybe,
reasoning by analogy, we will then be able to unravel the individual values of
$F_1$ and $F_2$.

At leading order, the effective QCD Lagrangian for two massless flavors
(up- and down-quark mass equal to zero) involves a single coupling constant,
$F_{\pi}$ (see e.g. {Leutwyler Brasil}),
\begin{equation}
{\cal L}^{QCD}_{eff} = \mbox{$ \frac{1}{2}$} F^2_{\pi} {\partial}_{\mu}U^i
{\partial}^{\mu}U^i ,
\end{equation}
whereas the effective Lagrangian of an antiferromagnet involves two such
quantities,
\begin{equation}
{\cal L}^{AF}_{eff} \, = \, \mbox{$ \frac{1}{2}$} F^2_1 {\partial}_0U^i
{\partial}_0U^i - \mbox{$ \frac{1}{2}$} F^2_2 {\partial}_rU^i {\partial}_rU^i
.
\end{equation}
As it is commonly done with the velocity of light in relativistic theories, we
may put the spin-wave velocity $v = {F_2}/{F_1}$ to one. In this "$\hbar = v =
1$"-system the two coupling constants coincide: $F_1 = F_2 \equiv F_{AF}$.
Also, the two Lagrangians above are then the same, except for the number of
fields $U^i$ and the actual values of the couplings $F_{\pi}$ and
$F_{AF}$. In this framework, where only one low energy constant
$F_{AF}$ exists, neutron scattering experiments again, of course, do not
shed any light on its value -- they merely fix the fundamental scale of the
spin-wave velocity $v$ in the respective crystal, analogous to a determination
of the velocity of light, which is then put to one.

Now, in the Standard Model of elementary particle physics, $F_{\pi}$ is
related to the electroweak interaction: the constant shows up in the
description of pion-decay processes and can be determined by measuring the
lifetime of charged pions -- $F_{\pi}$ is therefore referred to as pion decay
constant. This neat way of fixing $F_{\pi}$, offered by nature, has to be
regarded as a present from heaven, which, unfortunately, does not repeat itself
in an analogous manner for the antiferromagnet. Later on, in connection with
spin-wave scattering, we will take up the question of how to phenomenologically
determine $F_{AF}$ anew.

\section{Low Energy Theorem for Ferromagnets}
\label{LETFerro}

Let us now turn to our original intention, namely, to study the low energy
behavior of spin-wave scattering within the effective description. To begin
with, consider an elastic scattering process between two identical particles,
\begin{displaymath}
{\pi}({\vec k}_1) \, + \, {\pi}({\vec k}_2) \ \, \to \ \, {\pi}({\vec k}_3) \,
+ \, {\pi}({\vec k}_4) .
\end{displaymath}
In a nonrelativistic normalization of the one-magnon states,
\begin{equation}
\langle {\pi}(\vec k) \, | \, {\pi}(\vec k') \rangle \, = \, (2{\pi})^3 \,
{\delta}^3({\vec k} - {\vec k'}) ,
\end{equation}
the $S$-matrix relating to is given by
\begin{eqnarray}
\lefteqn{S \ = \ {\langle \pi({\vec k}_3) \, \pi({\vec k}_4) \, out \ | \
\pi({\vec k}_1) \, \pi({\vec k}_2) \, in \rangle}} \hspace{0.5cm}
\nonumber\\& & = \
(2 \pi)^6 \, \{ {\delta}^3({\vec k}_1 - {\vec k}_3) \, {\delta}^3 ({\vec k}_2 -
{\vec k}_4) \, + \, {\delta}^3({\vec k}_1 - {\vec k}_4){\delta}^3({\vec k}_2 -
{\vec k}_3) \} \nonumber\\& & \quad \mbox{} + \ i \, (2 \pi)^4 \,
{\delta}({\omega}_1 + {\omega}_2 - {\omega}_3 - {\omega}_4) \,
{\delta}^3({\vec k}_1 + {\vec k}_2 - {\vec k}_3 - {\vec k}_4) \; T .
\end{eqnarray}
As far as the evaluation of the $T$-matrix element is concerned, we will lean
on the canonical approach, since the calculation is more readily done by means
of field operators, rather than by making use of the external field technique.
Within the latter framework, where one uses the equation of motion to
evaluate appropriate four-point functions, the analysis of ferromagnetic
spin-wave scattering, although straightforward, turns out to be tedious. The
point is that, due to contributions proportional to $f^a_0 U^a$ appearing in
the effective Lagrangian, a second order iteration of the Landau-Lifshitz
equation is required -- accordingly, a careful bookkeeping is advised.
Nonetheless, the final result coincides with the one to be obtained below: of
course, it does not depend on the specific procedure used. Let us now briefly
provide ourselves with the tools needed in the canonical formalism -- we start
with the ferromagnet and construct the corresponding magnon field operators in
the interaction picture.

The Lagrangian is split up into two parts,
\begin{equation}
{\cal L} \ = \ {\cal L} \, {|_{f = 0}} \, + \, f^i_{\mu} J^{\mu}_i \, + \,
{\cal O} (f^2) . 
\end{equation}
Making use of the effective expression for ferromagnets, ${\cal L}
\Leftrightarrow {\cal L}^{F}_{eff}$ (\ref{LeffFerro}), the charge densities
are identified as
\begin{equation}
(J^0_i)_{eff} = \Sigma U^i .
\end{equation}
Recalling the transition matrix element (\ref{FerroMatrixTime}) relating to,
\begin{displaymath}
\langle 0|\, J^0_a(x) |\,{\pi}(\vec k) \rangle \, = \, {\varepsilon}_a
{\sqrt{\Sigma}} \, e^{-ikx} , \qquad {\varepsilon}_a =
\mbox{$ \frac{1}{\sqrt{2}}$} (1,-i) ,
\end{displaymath}
we finally arrive at the representation of the magnon field operators $U^a$ in
the interaction picture,
\begin{eqnarray}
U^a(x) \, = \, \frac{1}{\sqrt{\Sigma}} \int \! \! \frac{d^3k}{(2{\pi})^3} \ \{
{\varepsilon}_a \, a({\vec k} \,)e^{-ikx} \, + \, {\varepsilon}^{\star}_a \,
a(\vec k)^{\dagger} \, e^{ikx} \} , \qquad \\ \mbox{} [ a (\vec k),
a({\vec k'})^{\dagger} ] \ = \ (2 \pi)^3{\delta}^3({\vec k} - {\vec k'}) ,
\qquad | {\pi}(\vec k) \rangle \ = \ a({\vec k})^{\dagger} \,|0 \rangle
\nonumber .
\end{eqnarray}
The magnon field operators to be used below, read
\begin{equation}
\label{FieldOpFerro}
u(x) \, = \, \sqrt{\frac{2}{\Sigma}} \int \! \! \frac{d^3k}{(2\pi)^3} \, a(\vec
k) e^{-ikx} , \qquad u(x)^{\dagger} \, = \, \sqrt{\frac{2}{\Sigma}} \int \!
\! \frac{d^3k}{(2\pi)^3} \, a(\vec k)^{\dagger} e^{ikx} .
\end{equation}
Note that, on the classical level, these operators correspond to the following
linear combinations of the components of the magnon field ${\vec U}$: $u = U^1
+ iU^2$ and $u^{\star} = U^1 - iU^2$, respectively.

Next, we determine those terms in the effective Lagrangian, which are relevant
to the scattering process in question. Turning off the external fields
$f^i_{\mu}(x)$ in the original expression ${\cal L}^{F}_{eff}$
(\ref{LeffFerro}) altogether, we find
\begin{equation}
{\cal L}^{F}_{eff}{|_{f = 0}} \ = \ \frac{\Sigma}{1 + U^3} \,
{\varepsilon}_{ab} {\dot U}^aU^b \, - \, \mbox{$\frac{1}{2}$}
F^2 {\partial}_rU^i{\partial}_rU^i , \qquad {\varepsilon}_{ab} = -
{\varepsilon}_{ba} , \quad {\varepsilon}_{12} = 1 .
\end{equation}
Expanding the field $U^3$,
\begin{equation}
U^3 \, = \, (1 \, - \, U^a U^a \, )^{1/2} \; = \; 1 \, - \,
\mbox{$\frac{1}{2}$} U^aU^a - \ldots ,
\end{equation}
the terms quartic in $U^a$, which describe the spin-wave interaction, can be
read off,
\begin{equation}
{\cal L}^{F}_{int} \; = \; \mbox{$\frac{1}{8}$} \Sigma {\varepsilon}_{ab}
{\dot U}^aU^b(U^cU^c) \, - \, \mbox{$\frac{1}{2}$}F^2(U^a{\partial}_rU^a)(U^b
{\partial}_rU^b) .
\end{equation}
Written in terms of the field operators $u$ and $u^{\dagger}$, the relevant
expression is given by
\begin{equation}
\label{LeffFsymm}
{\cal L}^{F}_{int} \; = \; \mbox{$\frac{1}{16}$} i \Sigma \, (u^{\dagger}
u^{\dagger}u{\dot u} - uuu^{\dagger}{\dot u}^{\dagger}) \, - \,
\mbox{$\frac{1}{8}$} F^2 {\partial}_r(u^{\dagger} u)
{\partial}_r(u^{\dagger}u) .
\end{equation}
With the representation (\ref{FieldOpFerro}) of the field operators, the
evaluation of the $T$-matrix element is readily done, resulting in
\begin{equation}
\label{T-Ferro}
T^{F} \; = \; \langle \pi({\vec k}_3) \, \pi({\vec k}_4) \, |
\, {\cal L}^{F}_{int} \, | \, \pi({\vec k}_1) \, \pi({\vec k}_2) \rangle \; =
\; \frac{2\gamma}{\Sigma} \, {\vec k}_1 \! \cdot {\vec k}_2 .
\end{equation}
For the corresponding differential cross section, we obtain (see appendix B)
\begin{equation}
{\frac{d\sigma}{d\Omega}}^{F} \, = \, \frac{1}{32{\pi}^2{\Sigma}^2}
\, ({\vec k}_1 \! \cdot {\vec k}_2)^2 .
\end{equation}
The evaluation of the total cross section is trivial, because the $T$-matrix
element (\ref{T-Ferro}), remarkably, does not depend on any angles associated
with the outgoing particles. For ferromagnetic spin-wave scattering, the low
energy theorem for the total cross section thus amounts to
\begin{equation}
\label{DysonScatter}
{\bf \sigma}^{F}_{tot} \, = \, \frac{1}{8\pi{\Sigma}^2} \, ({\vec k}_1 \!
\cdot {\vec k}_2)^2 .
\end{equation}
This result is the same as the one Dyson derived in his microscopic theory of
spin waves a long time ago \cite{Dyson1}. Clearly, the expression obtained,
although invariant under space rotations, violates Lorentz symmetry -- a
peculiarity, that can only occur if the ground state of the theory is
Lorentz-noninvariant.

Whereas the above expression only reflects the (isotropic) S-wave part of the
scattering cross section, Dyson worked out all terms to the order considered.
In particular, the anisotropy of the lattice manifests itself in the
scattering reaction: for each one of the three types of cubic crystals, he
gets, in addition to the term (\ref{DysonScatter}), D-wave contributions. In
the framework of the effective expansion, these terms only show up at
next-to-leading order. Even with these additional contributions, the resulting
scattering amplitude would not be the whole story: since all these expressions
are real, the scattering amplitude does not satisfy the requirements imposed
by unitarity. If we had included loop corrections to our tree level
calculation, imaginary contributions in the scattering amplitude would then
have shown up.

An analogous feature arises in quantum chromodynamics, where the
next-to-leading order effective Lagrangian has been worked out already some
time ago \cite{Gasser Leutwyler QCD}. While with the leading order effective
QCD Lagrangian a concise rederivation of low energy theorems concerning the
pions may be achieved, the next-to-leading order Lagrangian as well as loop
graphs, originating from the leading order contribution, permit to
systematically correct these theorems. In particular, imaginary terms
resulting from loop graphs play a decisive role in the pion-pion scattering
amplitude, which has been worked out to even higher orders
\cite{Colangelo et al}. As it is characteristic of the effective Lagrangian
method, new effective coupling constants appear, if one extends $\chi$PT to
higher orders of momentum. As far as QCD is concerned,\cite{footnote6} two such
new couplings show up in the next-to-leading order effective Lagrangian, which
are left undetermined by chiral or Lorentz symmetry, and hence have to be fixed
phenomenologically. In fact, the analysis of pion-pion scattering experiments
leads to a determination of these fundamental constants of chiral perturbation
theory.

Unfortunately, we are not in an equally satisfactory position as far as
magnetic systems are concerned: in order to experimentally detect spin-wave
interactions, scattering processes are not the suitable tool -- the
corresponding cross section turns out to be very small (see e.g. \cite{Kittel
QT}). As far as I know, no experiments making this quantity directly
accessible, have ever been performed. Having this experimental situation in
mind, it would certainly not be a very clever idea to work out the effective
Lagrangian of a ferromagnet to next-to-leading order, with the only intention
to elaborate the analysis of spin-wave scattering further -- an experimental
determination of these additional effective couplings, appearing at higher
orders of momentum, clearly cannot come from this sector. Rather, the access
to some of these low energy constants will be made available by another field:
applications of the method to thermodynamic quantities, such as the variation
of the magnetization with temperature, may be of considerable help to carry
through this program.

In any case, at leading order of the derivative expansion, the effective
Lagrangian method reproduces the low energy theorem found by Dyson. Once the
machinery is developed, predictions for quantities of physical interest are
obtained in a concise and straightforward manner. In particular, for the
total cross section to exhibit the above Lorentz-noninvariant structure, it is
essential that the effective Lagrangian of a ferromagnet contains a
contribution, proportional to the spontaneous magnetization, which is not
invariant under the symmetry group G = O(3) -- from a methodical point of view,
this is probably the main conclusion to be drawn from this section.

\section{Low Energy Theorem for Antiferromagnets}
\label{LETAF}

Turning now to antiferromagnetic spin-wave scattering, we are faced with a
minor complication arising from the fact that there exist two independent
polarization states. Accordingly, the interaction in question may generally be
formulated as
\begin{displaymath}
{\pi}^a({\vec k}_1) \, + \, {\pi}^b({\vec k}_2) \ \to \ {\pi}^c({\vec k}_3) \,
+ \, {\pi}^d({\vec k}_4) , \qquad \quad a, \ldots, d \, = \, 1,2 .
\end{displaymath}
Analogous to the preceding paragraph, we are going to use a nonrelativistic
normalization of the one-magnon states,
\begin{equation}
\langle {\pi}^a(\vec k) \, | \, {\pi}^b(\vec k') \rangle \, = \, (2{\pi})^3 \,
{\delta}^{ab} {\delta}^3({\vec k} - {\vec k'}) ,
\end{equation}
and the evaluation of the $T$-matrix element will be based on the canonical
approach.

So again, the Lagrangian is split up into two parts,
\begin{displaymath}
{\cal L} \ = \ {\cal L} \, {|_{f = 0}} \, + \, f^j_{\mu} J^{\mu}_j \, + \, {\cal
O} (f^2) .
\end{displaymath}
With the effective expression for antiferromagnets, ${\cal L}
\Leftrightarrow {\cal L}^{AF}_{eff}$ (\ref{LeffAF}), the currents are
identified as
\begin{equation}
\label{Jreff}
(J^r_j)_{eff} \, = \, - F^2_2 {\varepsilon}_{ijk} {\partial}_rU^iU^k .
\end{equation}
Considering the transition matrix element (\ref{AFmatrix}) relating to,
\begin{displaymath}
\langle 0|\, J^r_a(x) |{\pi}^b(\vec k) \rangle \, = \, i {\delta}^b_a k^r v F_2
e^{-ikx} / \sqrt{2\omega} ,
\end{displaymath}
the magnon field operators, associated with the two polarization states,
read:\cite{footnote7}
\begin{eqnarray}
U^a(x) \, = \, \frac{v}{F_2} \int \! \!
\frac{d^3k}{(2\pi)^3 \sqrt{2 \omega}} \, \{ a^a({\vec k})e^{-ikx} +
a^a({\vec k})^{\dagger} e^{ikx} \} , \qquad \\
\mbox{} [ a^a (\vec k), a^b(\vec k')^{\dagger} ] \, = \, (2\pi)^3 \,
{\delta}^{ab} {\delta}^3 ({\vec k} - {\vec k'}) , \qquad |{\pi}^a (\vec k)
\rangle = {\varepsilon}_{ab} a^b(\vec k)^{\dagger}\,|0 \rangle \nonumber .
\end{eqnarray}
In order to determine the relevant interaction terms in the effective
Lagrangian ${\cal L}^{AF}_{eff}$ (\ref{LeffAF}), we put the external fields
$f^i_{\mu}(x)$ to zero,
\begin{equation}
{\cal L}^{AF}_{eff}{|_{f = 0}} \, = \, \mbox{$\frac{1}{2}$} F^2_1
{\partial}_0U^i {\partial}_0U^i - \mbox{$\frac{1}{2}$} F^2_2 {\partial}_r
U^i {\partial}_rU^i ,
\end{equation}
expand the variable $U^3$, and extract the terms quartic in $U^a$,
\begin{equation}
\label{LeffAFsymm}
{\cal L}^{AF}_{int} \, = \, \mbox{$\frac{1}{2}$} F^2_1 (U^a{\partial}_0U^a)
(U^b {\partial}_0U^b) - \mbox{$\frac{1}{2}$} F^2_2 (U^a{\partial}_rU^a)
(U^b{\partial}_rU^b) .
\end{equation}
We then obtain the following low energy theorem for the T-matrix element
describing antiferromagnetic spin-wave scattering:
\begin{eqnarray}
\label{T-AFiso}
\lefteqn{T^{AF} \ = \ \langle {\pi}^c({\vec k}_3) \, {\pi}^d({\vec k}_4) \, |
\; {\cal L}^{AF}_{int} \; | \, {\pi}^a({\vec k}_1) \, {\pi}^b({\vec k}_2) 
\rangle} \nonumber\\& & \quad =
\frac{1}{2\sqrt{{\omega}_1{\omega}_2{\omega}_3{\omega}_4}}
\frac{v^4}{F^2_2} \ \{ \, {\delta}^{ab} {\delta}^{cd} \, (|{\vec k}_1||{\vec
k}_2| - {\vec k}_1 \! \cdot {\vec k}_2) \ - \ {\delta}^{ac}{\delta}^{bd} \,
(|{\vec k}_1||{\vec k}_3| - {\vec k}_1 \! \cdot {\vec k}_3) \hspace{0.7cm}
\nonumber\\& & \mbox{} \qquad \qquad \qquad \qquad \qquad - \
{\delta}^{ad}{\delta}^{bc} \, (|{\vec k}_1| |{\vec k}_4| - {\vec k}_1 \! \cdot
{\vec k}_4) \, \} .
\end{eqnarray}
As far as actual measurements of the scattering cross section relating to are
concerned, we are in an equally unsatisfactory position as that of the
ferromagnet before. This experimental dead end is indeed highly unwelcome,
since we have not yet been able to determine the individual values of the two
low energy constants $F_1$ and $F_2$, occurring in the leading order effective
Lagrangian of the antiferromagnet. From the dispersion law we merely know their
ratio, $v = F_2/F_1$. Note that the constant $F_2$ shows up separately in
formula ({\ref{T-AFiso}). In principle then, a measurement of the corresponding
cross section would offer the possibility to fix this constant and hence allow
to extract the other coupling $F_1$ from experimental data.

Turning now to the theoretical side, the literature on antiferromagnetic
spin-wave scattering appears to be rather scarce. Unlike for the ferromagnet,
where, after Dyson's monumental work, a whole lot of publications on the
subject showed up (some of them trying to simplify his calculations and
rederive his results, see e.g. \cite{Keffer Loudon}), only a few references
dealing with the analogous problem in antiferromagnets seem to be available.
References \cite{Oguchi,Beeman Pincus,Brooks Harris} rely on a microscopic
description of the antiferromagnet, while reference \cite{Kashcheev}
approaches the subject on the basis of a phenomenological theory. However,
these authors rather direct their attention to other aspects of the spin-wave
interaction. Moreover, the paper of Brooks Harris \cite{Brooks Harris} appears
to be the only one which is in agreement with the energy-momentum dependence
of the scattering amplitude (\ref{T-AFiso}) obtained above.

From a methodical point of view, it is instructive to compare the result
regarding antiferromagnetic spin-wave scattering with what is known about the
analogous item in QCD: pion-pion scattering. There, at leading order of the
effective expansion, the $T$-matrix element in question takes the
Lorentz-invariant form\cite{footnote8} (see e.g. \cite{Gasser Leutwyler QCD})
\begin{equation}
\label{T-QCD}
T^{QCD} \ = \ \frac{1}{F_{\pi}^2} \ \{ \, {\delta}^{ab} {\delta}^{cd}
\, s \; + \; {\delta}^{ac} {\delta}^{bd} \, t \; + \; {\delta}^{ad}
{\delta}^{bc} \; u \, \} ,
\end{equation}
where $s, t, u$ denote the Mandelstam variables,
\begin{equation}
s = (k_1 + k_2)^2 , \quad t = (k_1 - k_3)^2 , \quad u = (k_1 - k_4)^2 .
\end{equation}
The indices $a, \ldots ,d$ in (\ref{T-QCD}), labeling the three different
isospin states, are analogous to the ones needed to denote the two independent
polarization states of antiferromagnetic magnons in (\ref{T-AFiso}). Comparing
these two amplitudes, we see that the energy-momentum dependence is the same:
the respective terms in (\ref{T-AFiso}) may be viewed as scalar products of
momentum four vectors. As a matter of fact, the analogy between the two
expressions is even more pronounced. If, just for the moment, a relativistic
normalization of the one-magnon states is used and the spin-wave velocity $v$
is put to one, "$\hbar=v=1$" $\, \to F_1 = F_2 \equiv F_{AF}$, then they
formally coincide: apart from the number of independent Goldstone states and
the actual values of the constants $F_{\pi}$ and $F_{AF}$, the two
formulas are identical.

Clearly, this finding does not come about unexpectedly. The similarity between
the effective Lagrangian of an antiferromagnet and the one of QCD is
transferred to the scattering amplitudes: they exhibit analogous low energy
representations. In fact, the above example may serve as a nice illustration
of a characteristic feature of the effective Lagrangian technique --
universality. Let us close this paragraph with some remarks on the subject.

In the construction of the effective Lagrangian, the specific properties of
the underlying theory do not matter: they merely affect the numerical values of
the coupling constants appearing in ${\cal L}_{eff}$. The only relevant
information is the structure of the two groups G and H, associated with the
exact symmetry of the underlying theory -- the low energy description turns out
to be universal. Now, QCD with two massless flavors displays an exact
$\mbox{SU(2)}_{R} \times \mbox{SU(2)}_{L}$ symmetry which is
spontaneously broken to $\mbox{SU(2)}_{V}$; these two groups are locally
isomorphic to G = O(4) and H = O(3), respectively. Hence, the analogy to the
\{O(3) $\to$ O(2)\}-antiferromagnet, considered in this paper, is almost
perfect: except for the magnitude of the constants $F_{AF}$ and $F_{\pi}$,
the two effective Lagrangians also differ in the number of Goldstone particles.

Since ferro- and antiferromagnets, in our approach, are undistinguishable from
the point of view of symmetry, these two nonrelativistic systems should
actually provide us with a perfect illustration of the universality concept. It
so happens, however, that, for the latter system, one of the low energy
constants, the spontaneous magnetization, turns out to be {\it zero}. As a
consequence, the effective Lagrangians relating to are apparently different,
although, in either case, their construction is based on the symmetry groups G
= O(3) and H = O(2) inherent in the Heisenberg model. Note that, nonetheless,
the concept of universality applies -- the specific properties of an
antiferromagnet, however, manifest themselves in a rather drastic way.

This striking difference in the structure of these two Lagrangians on the
effective level is quite remarkable, because, in the underlying theory, the
respective Hamiltonians only differ in the sign of the exchange integral
$J$. A contragredient behavior, now really illuminating the concept of
universality, concerns the antiferromagnet and the strong interaction: although
the underlying theories, the Heisenberg model and QCD, respectively, are
completely different, the corresponding effective Lagrangians are almost the
same.

\section{External Magnetic Field}
\label{ExSB}

Up to now, the analysis of spontaneous symmetry breaking was related to {\it
exact} symmetries: it was assumed that the underlying theory is invariant with
respect to an internal symmetry group G. In what follows in the remaining part
of this presentation, we will let aside this idealization and direct our
attention to {\it approximate} symmetries. The low energy phenomena considered
so far will be studied in this extended framework and their modification,
imposed by explicit symmetry breaking, will be discussed.\cite{footnote9}

As a first example of explicit symmetry breaking, let us work out the effect of
an external magnetic field on the low energy behavior of ferro- and
antiferromagnets. On a microscopic level, the interaction between a constant
magnetic field ${\vec H}$ and the spin degrees of freedom is taken into account
through the Zeeman term. In the corresponding extension of the Heisenberg
model,\cite{footnote10}
\begin{equation}
\label{Zeeman}
{\cal H} \ = \ {\cal H}_0 \, - \, {\mu}{\sum_n}{\vec S}_n \! \cdot {\vec H} ,
\end{equation}
the magnetic field is coupled to the vector of the total spin. Whereas ${\cal
H}_0$ is invariant under a simultaneous rotation of the spin variables, the
second term explicitly breaks the symmetry with respect to the group G = O(3).
In the effective Lagrangian framework, the interaction with a magnetic field
corresponds to the term $\int \!\! d^3x f^i_0 J^0_i$: the operator of the total
spin, ${\sum}_n{\vec S}_n$, is to be identified with $\int \!\! d^3x
{\vec J}^0$, while the magnetic field ${\vec H}$, playing the role of a
symmetry breaking parameter, is related to the time components of the auxiliary
field, $f^i_0 = \mu H^i$.

Independently of whether the effective description refers to a ferro- or an
antiferromagnet, an external magnetic field is taken into account through the
quantities $f^i_0(x)$. Apart from the identification $f^i_0 = \mu H^i$,
nothing further has to be done -- the effective machinery developed earlier
applies as it stands. However, in order to obtain the change in low energy
structure induced by the magnetic field, the effective expansion is to be
performed around the nonzero, constant value of $\mu H^i$ appearing in
the underlying theory, i.e. in the extended Heisenberg model (\ref{Zeeman}).

As far as the ferromagnet is concerned, the magnetic field ${\vec H} = (0, 0,
H) \, , H > 0$, couples to the order parameter: it enters the leading order
effective Lagrangian (\ref{LeffFerro}) through a term proportional to the
spontaneous magnetization,
\begin{displaymath}
{\cal L}^{F}_{eff} ({\vec H}) \ = \ \Sigma \, \frac{\varepsilon_{ab}
{\partial}_0 U^a U^b}{1 + U^3} \, + \, \Sigma \mu H^i U^i \, - \,
\mbox{$ \frac{1}{2}$} F^2 {\partial}_rU^i{\partial}_rU^i .
\end{displaymath}
Expanding $U^3 = (1 - U^aU^a)^{1/2}$ in powers of the two components $U^a, \; a
= 1,2$, the term in question gives rise to the following contributions:
\begin{equation}
\label{LeffFerroH}
\Sigma \mu H^i U^i \, = \, \Sigma
\mu H \, (1 - \mbox{$ \frac{1}{2}$} U^aU^a - \mbox{$ \frac{1}{8}$} U^aU^a
U^bU^b - \ldots \ ) .
\end{equation}
The linearized equation of motion shows that, in the presence of an external
magnetic field, the dispersion law of ferromagnetic spin waves keeps its
quadratic structure, the corresponding coefficient $\gamma$ being unchanged.
The energy of the single spin-wave branch, $u = U^1 + iU^2$, is merely shifted
by a constant amount, proportional to the symmetry breaking parameter,
\begin{equation}
\label{disprelFH}
\omega \, = \, \gamma {\vec k}^2 + \mu H .
\end{equation}
Much like an approximate chiral symmetry provides the pions with a mass, an
approximate symmetry with respect to internal rotations, G = O(3), causes an
energy gap in the spin-wave spectrum of a ferromagnet, $\Delta \omega = \mu
H$. Note that the spontaneous magnetization drops out in this expression: the
energy gap is determined by the measure of {\it explicit} symmetry breaking
alone.

Due to the term quartic in the magnon variables in (\ref{LeffFerroH}), the
scattering amplitude of ferromagnetic spin waves seems to experience a
modification by the magnetic field as well. However, the resulting extra term
in the T-matrix element is canceled by the contribution originating from the
unperturbed Lagrangian (\ref{LeffFsymm}), evaluated with the dispersion
relation (\ref{disprelFH}) -- hence, to the order considered, the interaction
in question is not affected by a magnetic field.

As far as the antiferromagnet is concerned, an external magnetic field does not
manifest itself in an analogous manner in the low energy expansion: terms
involving the spontaneous magnetization do not occur in the effective
Lagrangian (\ref{LeffAF}). Rather, the magnetic field appears in the time
component of the covariant derivative of ${\vec U}$,
\begin{eqnarray}
{\cal L}^{AF}_{eff} ({\vec H}) \, = \, \mbox{$ \frac{1}{2}$} F^2_1
D_0U^iD_0 U^i - \mbox{$ \frac{1}{2}$} F^2_2 {\partial}_rU^i{\partial}_rU^i
, \nonumber\\D_0U^i \, = \, {\partial}_0 U^i + {\varepsilon}_{ijk} \mu H^j U^k
\nonumber .
\end{eqnarray}
Concentrating on those contributions which involve the magnetic field, the
expansion yields
\begin{equation}
\label{LeffAFH}
F^2_1 \mu H \, \{ \, - {\varepsilon}_{ab} {\partial}_0 U^aU^b + \mbox{$
\frac{1}{2}$} \mu H U^aU^a \} .
\end{equation}
The linearized equation of motion leads to the dispersion relation
\begin{equation}
\label{disprelAFH}
\omega_{\pm} \ = \ v|{\vec k}| \, \pm \mu H .
\end{equation}
In the presence of ${\vec H}$, the dispersion law of antiferromagnetic spin
waves keeps its linear structure -- as for a ferromagnet before, it is merely
shifted by a constant amount, proportional to the symmetry breaking field. Note
that the two independent spin-wave branches, $u = U^1 + iU^2$ and $u^{\star} =
U^1 - iU^2$, respectively, are affected in distinct ways: the magnetic field
lifts their degeneracy by splitting them up symmetrically. Remarkably, the
magnetic field does not give rise to a "mass term": in the case of a
relativistic dispersion relation, as we see it here with the antiferromagnet,
such a term would show up under a square root,
\begin{equation}
\label{AFmass}
\omega = \sqrt{v^2 {\vec k}^2 + v^4 M^2_{GB}} .
\end{equation}
Finally, let us consider the effect of an external magnetic field on
antiferromagnetic spin-wave scattering. Remarkably, the expansion
(\ref{LeffAFH}) does not contain any terms quartic in the magnon fields. Now,
in order to evaluate the T-matrix element referring to the unperturbed
effective Lagrangian (\ref{LeffAFsymm}) with the dispersion relation
(\ref{disprelAFH}), we have to choose the representation of the two
polarization states accordingly: $|{\pi}^{+}\rangle$ ($|{\pi}^{-}\rangle$)
corresponds to the spin-wave branch $u = U^1 + iU^2$ ($u^{\star} = U^1 -
iU^2$), which experiences a positive (negative) shift by ${\vec H}$. The
calculation shows that the respective T-matrix elements do not receive
additional terms from the magnetic field.

Take for example the reaction ${\pi}^{+}({\vec k}_1) \, + \, {\pi}^{-}({\vec
k}_2) \ \to \ {\pi}^{-}({\vec k}_3) \, + \, {\pi}^{+}({\vec k}_4)$, which
yields
\begin{eqnarray}
\lefteqn{T^{AF}(\vec H) \ = \ \langle {\pi}^{-}({\vec k}_3) \, {\pi}^{+}({\vec
k}_4) \, | \; {\cal L}^{AF}_{int} \; | \, {\pi}^{+}({\vec k}_1) \,
{\pi}^{-}({\vec k}_2) \rangle} \nonumber\\& & \qquad \quad = \
\frac{1}{4\sqrt{{\omega}_1{\omega}_2{\omega}_3{\omega}_4}} \frac{v^2}{F^2_2} \
\{ \, ({\omega}_1 + {\omega}_3)^2 - v^2 ({\vec k}_1 + {\vec k}_3)^2 \} .
\end{eqnarray}
The magnetic field drops out in the sum ${\omega}_1 + {\omega}_3$ -- it only
appears in the denominator of the scattering amplitude, which exhibits the
dispersion relation (\ref{disprelAFH}).

\section{Anisotropy Field}
\label{AnisoField}

While the preceding section was devoted to a single symmetry breaking
parameter, an external magnetic field, we would now like to discuss the
question of explicit symmetry breaking from a general point of view. In the
case of an approximate symmetry, the Lagrangian of the underlying theory
contains contributions, which explicitly break the internal symmetry
associated with the group G,
\begin{equation}
{\cal L} \ = \ {\cal L}_0 \, + \ m_{\alpha} O^{\alpha} .
\end{equation}
Whereas the first term represents the invariant part, the operators
$O^{\alpha}$ transform nontrivially under the symmetry group G. The constants
$m_{\alpha}$, for their part, play the role of symmetry breaking parameters.

In this perspective, the interaction of an external magnetic field with the
spin degrees of freedom represents a special case: the operators $O^{\alpha}$
are to be identified with the charge densities $J^0_i$, and are thus related
to the generators $Q_i$ of the group G. Hence, in the effective description,
the symmetry breaking parameters $m_{\alpha}$ of the underlying theory are to
be taken into account through the time components of the external field
$f^i_0(x)$ -- in connection with explicit symmetry breaking, these auxiliary
fields, as we have seen before, acquire physical significance.

If the operators $O^{\alpha}$ are not related to the generators of the group
G, then the effective Lagrangian has to be enlarged, including additional
contributions which take into account the approximate character of the
spontaneously broken symmetry. It is convenient to extend the effective
machinery accordingly, treating the corresponding symmetry breaking parameters
$m_{\alpha}$ also as external fields, $m_{\alpha}(x)$, on the same footing as
the vector fields, $f^i_{\mu}(x)$, associated with the currents and charge
densities. The generating functional then contains two arguments, $\Gamma =
\Gamma \{ f, m \}$. Correlation functions of the novel operators $O^{\alpha}$
may be obtained the same way as those involving the currents and charge
densities. The only modification brought about by the fields $m_{\alpha}(x)$
is that the low energy expansion of the functional $\Gamma \{f, m \}$ now
amounts to a double series -- in an expansion in powers of the external fields
$f^i_{\mu}(x)$, as well as in an expansion in powers of the quantities
$m_{\alpha}(x)$, associated with the novel operators $O^{\alpha}$. As a
prototype of this more general way of symmetry breaking, we mention the quark
mass term, ${\bar q}{\cal M}q$, of the QCD Lagrangian: if the quark masses,
playing the role of symmetry breaking parameters, are taken at their physical,
nonzero values, chiral symmetry is explicitly broken. An application in
connection with nonrelativistic systems will be given below.

From a methodical point of view, the following observation related to order
parameters is of interest. Concerning the nature of these quantities,
nonrelativistic kinematics, as we have seen, is less restrictive: in the case
of a Lorentz-noninvariant ground state, the time components of conserved
currents may develop such nonzero vacuum expectation values. Now, this type of
order parameter, $\langle 0|\, J^0_i \,|0 \rangle$, already shows up in the
effective theory, if the underlying theory is {\it symmetric} -- the ferromagnet
represents such a system, where the spontaneous magnetization embodies this
possibility. Similarly, the vacuum expectation values of the more general
operators $O^{\alpha}$, which are not related to the generators of the group
G, also represent order parameters, which, for their part, may occur both in
Lorentz-invariant theories and in the nonrelativistic domain. However, this
type of order parameter only shows up in the effective theory, if the symmetry
of the underlying theory is {\it approximate} -- on the effective level, the
quantities $\langle 0|\, O^{\alpha} \,|0 \rangle$ then appear in association
with the fields $m_{\alpha}(x)$. As an illustration, referring to the
relativistic domain, we quote QCD, where the quark condensate,
$\langle 0|\, {\bar q}q \,|0 \rangle$, represents the order parameter in
question. Likewise, the staggered magnetization ${\Sigma}_s$ of an
antiferromagnet may serve as an example of an order parameter relevant to
condensed matter physics, which belongs to this more general class.

In what follows, we are going to consider a further extension of the
Heisenberg model,
\begin{equation}
\label{Zeeman-h}
{\cal H} \ = \ {\cal H}_0 \; - \; {\mu}{\sum_n} {\vec S}_n \! \cdot {\vec H}
\; - \; {\mu} {\sum_n (-1)^n} {\vec S}_n \! \cdot {\vec h} ,
\end{equation}
which illustrates the concept of explicit symmetry breaking exposed above. The
two fields, ${\vec H}$ and ${\vec h}$, are assumed to be weak, such that the
respective interaction terms involving the spin degrees of freedom, may be
considered as a perturbation of the isotropic Heisenberg Hamiltonian ${\cal
H}_0$. In the Zeeman term, the sum over the spin operators is associated with
the spontaneous magnetization, ${\vec {\Sigma}} \propto \langle 0|\, {\sum}_n
{\vec S}_n \,|0 \rangle $, while the vacuum expectation value of the second sum
is related to the staggered magnetization, ${\vec {\Sigma}_s} \propto \langle
0|\, {\sum}_n (-1)^n {\vec S}_n \,|0 \rangle$.

The field ${\vec h}$ in (\ref{Zeeman-h}) corresponds to those symmetry
breaking parameters $m_{\alpha}$, which are not associated with the generators
of the group G. Since the quantity ${\vec h}$, much like a magnetic field,
transforms with the vector representation of G = O(3), the corresponding
additional contribution in the leading order effective Lagrangian exhibits the
same structure as the effective representation of the Zeeman term
(\ref{LeffFerroH}),
\begin{equation}
\label{LeffHh}
{\cal L}_{eff}({\vec h}) \; = \; {\Sigma}_s \mu h^i U^i .
\end{equation}
The staggered magnetization, ${\Sigma}_s$, enters the leading order Lagrangian
in the form of a new coupling constant, whose value has yet to be determined.

As we have seen earlier, a term proportional to the spontaneous magnetization
does not appear in the effective Lagrangian of the antiferromagnet. For
ferromagnets, however, the contribution (\ref{LeffFerroH}) is substantial: it
describes the modification of the low energy structure imposed by a magnetic
field. On the other hand, the term (\ref{LeffHh}), which is proportional to
the staggered magnetization, does not show up in the effective Lagrangian of
the ferromagnet considered in this paper: in the case of identical spins at
each lattice site, we have ${\Sigma}_s = 0$. However, for antiferromagnets, it
is the staggered magnetization which is nonzero. Much like the auxiliary field
$f^i_0(x)$ acquires physical significance through a magnetic field ${\vec H}$,
the quantity $h^i$ in (\ref{LeffHh}) is related to a so-called anisotropy field
${\vec h}_{A}$. In any real magnetic system, there exist interactions whose
description is beyond the reach of the isotropic Heisenberg Hamiltonian. One
of these is magnetic anisotropy, which either originates from
dipol-dipol-interactions between the spins or may be caused by the coupling of
the electron orbits to the crystal field \cite{Nolting zwei}. In order to take
these interactions into account on a microscopic level, one may introduce the
artifice of an effective anisotropy field ${\vec h}_{A}$ into the
microscopic Hamiltonian. For the antiferromagnet, which is then referred to as
uniaxial, one obtains (see e.g. \cite{Kittel QT,van Kranendonk van Vleck})
\begin{eqnarray}
\label{ZeemanAniso}
{\cal H}^{AF} \; = \; -J \, \sum_{n.n.} {\vec S}_m \cdot
{\vec S}_n \, - \, \mu \! \! \sum_{n_a n_b} ({S}^3_{n_a} + {S}^3_{n_b}) \!
\cdot \! {H} \, - \, \mu \! \!\sum_{n_a n_b} ({S}^3_{n_a} - {S}^3_{n_b}) \!
\cdot \! {h_{A}} .
\end{eqnarray}
In this model, which represents a special case of the Hamiltonian
(\ref{Zeeman-h}), the antiferromagnet is considered as composed of two
sublattices $a$ and $b$, where $a$- and $b$-spins are of equal magnitude. The
arrangement is such that all nearest neighbors of an $a$-spin are $b$-spins
and vice versa. In an idealized picture of the ground state, $a$-spins point
up and $b$-spins point down. Note that, unlike for the external magnetic field
before, we are now dealing with a hypothetical field, which changes its
direction over atomic distances: ${\vec h}_{A}$ points along the positive
$3^{rd}$ axis at $a$-sites, but along the negative $3^{rd}$ axis at $b$-sites.

As far as ferromagnets are concerned, magnetic anisotropies manifest themselves
in a different manner in the microscopic description: again, they may be taken
into account through an effective field, ${\vec H}_{A}$, which locally
points in the same direction as every single spin vector; but here, they all
point along one and the same direction. Accordingly, this field ${\vec
H}_{A}$, which is also referred to as anisotropy field (see e.g. \cite{van
Kranendonk van Vleck,Keffer,Tjablikow,Herpin}), enters the microscopic
Hamiltonian through the term $- \mu {\sum}_n {\vec S}_n \! \cdot {\vec
H}_{A}$, i.e. it couples to the vector of the total spin, much like a
magnetic field ${\vec H}$. Therefore, on the effective level, it is also to be
incorporated into the quantities $f^i_0(x)$, $f^i_0(x) \Leftrightarrow \mu
H^i_{A}$, such that the qualitative effects of an external magnetic and an
anisotropy field, respectively, are the same.

For the antiferromagnet, on the other hand, anisotropy field ${\vec h}_{A}$
and magnetic field ${\vec H}$ are not to be treated in analogous ways, since
the quantity ${\vec h}_{A}$ does not couple to the vector of the total
spin. It is instructive to discuss this novel interaction in our effective
framework and confront the resulting modification of our previous findings with
what is known in condensed matter physics.

Let us first examine the spin-wave dispersion relation. According to the
preceding section, a magnetic field lifts the degeneracy of the two
polarization states observed in an antiferromagnet, but does not provide the
magnons with a "mass" -- an anisotropy field ${\vec h}_{A}$, however, does
the job. In the corresponding effective expansion,
\begin{equation}
\label{LeffAFaniso}
{\Sigma}_s \mu h^i_{A} U^i = \Sigma_s \mu h_{A} \, ( \, 1 - \mbox{$
\frac{1}{2}$} U^aU^a - \mbox{$ \frac{1}{8}$} U^aU^aU^bU^b - \ldots \ ) ,
\end{equation}
the term quadratic in the magnon variables leads to the relativistic scenario
referred to in (\ref{AFmass}): in the presence of an anisotropy field, ${\vec
h}_{A} = (0,0,h_{A})$, as well as of an external magnetic field,
${\vec H} = (0,0,H)$, the dispersion law of antiferromagnetic spin waves takes
the form\cite{footnote11}
\begin{equation}
\label{disprelAFanisoH}
\omega_{\pm} \, = \, \sqrt{v^2{\vec k}^2 + {\Sigma}_s \mu h_{A} / F^2_1}
\ \pm \, \mu H .
\end{equation}
Accordingly, the following relation holds,
\begin{equation}
\label{GOR-AF}
F^2_1 \, (v^2 M_{GB})^2 \, = \, {\Sigma}_s \, \mu h_{A} ,
\end{equation}
showing that the square of the "magnon mass" is proportional to the product of
order parameter, ${\Sigma}_s$, and symmetry breaking parameter, $\mu
h_{A}$. This formula may be viewed as the antiferromagnetic analog of the
well-known Gell-Mann--Oakes--Renner relation encountered in QCD
\cite{Gell-Mann Oakes Renner},
\begin{equation}
F^2_{\pi} M^2_{\pi} = | \langle 0|\, {\bar u}u \,|0 \rangle | \, (m_u + m_d) .
\end{equation}
The two equations are indeed in one-to-one correspondence: the square of the
pion mass is determined by the product of the order parameter, the quark
condensate $\langle 0|\, {\bar u}u \,|0 \rangle$, with the symmetry breaking
parameter, the sum of the quark masses $m_u + m_d$. While the first factor is a
measure of {\it spontaneous} symmetry breaking, the second one is a measure of
{\it explicit} symmetry breaking.

Finally, let us consider how the scattering amplitude of antiferromagnetic
spin waves is affected by the anisotropy field. From the expansion
(\ref{LeffAFaniso}), we derive the following low energy theorem:
\begin{eqnarray}
\label{T-AFisoh}
\lefteqn{T^{AF} ({\vec h}_{A}) \, = \, \frac{1}{4 \sqrt{{\omega}_1
{\omega}_2 {\omega}_3 {\omega}_4}} \, \frac{v^4}{F^2_2} \, \{ \,
{\delta}^{ab}{\delta}^{cd} \, ( \frac{2}{v^2} \; \! {\omega}_1 {\omega}_2 - 2
{\vec k}_1 \!\! \cdot \! {\vec k}_2 + {\Sigma}_s {\mu} h_{A} / {F^2_2})}
\hspace{3cm} \nonumber\\& & \mbox{} + {\delta}^{ac} {\delta}^{bd} \,
(2 {\vec k}_1 \!\! \cdot \! {\vec k}_3 - \frac{2}{v^2} \; \! {\omega}_1
{\omega}_3 + {\Sigma}_s \mu h_{A} / {F^2_2}) \nonumber\\& & \mbox{} +
{\delta}^{ad} {\delta}^{bc} \, (2 {\vec k}_1 \!\! \cdot \! {\vec k}_4 -
\frac{2}{v^2} \; \! {\omega}_1 {\omega}_4 + {\Sigma}_s \mu h_{A} / {F^2_2})
\, \} ,
\end{eqnarray}
where $\omega$ represents the modified dispersion relation, $\omega =
\sqrt{v^2 {\vec k}^2 + {\Sigma}_s \mu h_{A} / F^2_1}$. Again, the formula
(\ref{T-AFisoh}) has its counterpart in QCD, if the pion-pion scattering
amplitude is evaluated around nonzero quark mass.

In summary, the anisotropy field ${\vec h}_{A}$ is on the same footing as
the quark masses -- both quantities belong to those symmetry breaking
parameters $m_{\alpha}$, which are not related to the generators of the
symmetry group G. As we have pointed out in the analysis of symmetric
underlying theories, the leading order effective Lagrangian of an
antiferromagnet closely resembles the one describing QCD. Now, if the
effective framework is extended to approximate symmetries, including an
anisotropy field and quark masses, respectively, then the corresponding
analogy in the low energy structure of the two theories is maintained.

\section{Symmetry Breaking Parameters}
\label{SBP}

We have to recall that the entire analysis in the last two sections, concerning
approximate symmetries, relies on an essential assumption: the respective
contributions, which explicitly break the symmetry of the underlying theory,
are to be regarded as {\it perturbations} -- the analysis in terms of
effective fields is useful only, if the corresponding symmetry breaking
parameters are sufficiently small. Let us now focus on this important
requirement and discuss the various symmetry breaking parameters encountered
so far from this point of view. We start with the anisotropy field.

Assuming that this field is weak, the predictions of the effective Lagrangian
method, given in the previous section, can be trusted. For ferromagnets, as we
have seen, magnetic anisotropies may be taken into account through an effective
field ${\vec H}_{A}$, which is to be treated in the same way as a magnetic
field ${\vec H}$: the dispersion law of ferromagnetic spin waves experiences an
overall shift {\it linear} in the perturbation,
\begin{equation}
\omega = \gamma {\vec k}^2 + \mu H_{A} .
\end{equation}
For antiferromagnets, where the anisotropy field, ${\vec h}_{A}$, does not
couple to the generators of the group O(3), the perturbation shows up under a
{\it square root}, the corresponding coefficient being proportional to the
staggered magnetization,
\begin{equation}
\label{disprelAFaniso}
\omega = \sqrt{v^2 {\vec k}^2 + {\Sigma}_s \mu h_{A} / F^2_1} .
\end{equation}
Hence, if the anisotropy fields, ${\vec H}_{A}$ and ${\vec h}_{A}$,
respectively, are of the same order of magnitude and weak, the dispersion
relation of antiferromagnetic spin waves exhibits a larger energy gap.

Indeed, this striking difference concerning the significance of anisotropy
effects in ferro- and antiferromagnets is well-known in condensed matter
physics. In ferromagnets, these interactions only play a minor role, whereas
in antiferromagnets they are much more pronounced: as a microscopic analysis,
relying on some rough approximations, indicates (see e.g. \cite{Wagner}), the
spin-wave spectrum of an antiferromagnet exhibits a characteristic energy gap,
\begin{equation}
\label{gapAF}
\Delta \omega \, = \, \mu \sqrt{h_{A}(2h_{W} + h_{A})} ,
\qquad \quad {\vec k} \to 0 , \ \ {\vec H} \to 0 .
\end{equation}
$h_{W}$ is the so-called Weiss field, which turns out to be very large
compared to the anisotropy field, $h_{W} / h_{A} \approx 10^3$, such
that the second term can be neglected. Accordingly, the above combination of
anisotropy field and Weiss field, which does not show up in the analysis of
ferromagnets, may lead to a substantial energy gap in the spin-wave spectrum of
an antiferromagnet.

In particular, the formula for the energy gap (\ref{gapAF}) is consistent
with the dispersion law (\ref{disprelAFaniso}): the contribution involving
the Weiss field $h_{W}$ corresponds to the term involving the staggered
magnetization ${\Sigma}_s$ -- one identifies $h_{W} \Leftrightarrow
{\Sigma}_s / 2 \mu F^2_1$. The other term appearing under the square root in
(\ref{gapAF}), ${\mu}^2 h^2_{A}$, is not reproduced by the effective
theory: in our counting scheme, this expression corresponds to a contribution
of subleading order and is thus beyond the reach of the leading order effective
Lagrangian. Generally, our bookkeeping is based on a systematic counting of
powers of momentum, such that the respective terms of a given order need not
be correlated one-to-one with those obtained from a microscopic investigation
of condensed matter.

In the case of very strong anisotropy effects, antiferromagnetic spin waves no
longer follow the dispersion relation (\ref{disprelAFaniso}); rather, they
obey a quadratic law\cite{Nolting zwei}, $\omega = \alpha + \beta {\vec k}^2$ 
-- the effective description no longer applies. Chiral perturbation theory,
which is based on the assumption that the energy gap, associated with the
Goldstone bosons, is small, now breaks down. The fact that the effective
machinery only makes sense if the anisotropy field is weak, thus restricts the
range of application of the method. However, as far as condensed matter systems
are concerned, one has the freedom to chose appropriate objects of investigation
-- one may easily find another antiferromagnetic body, displaying a weaker
anisotropy field, and hence a smaller energy gap, such that the effective method
now perfectly applies.

Note the difference with the description of the strong interaction: QCD is a
universal theory -- the symmetry breaking parameters, the quark masses, are
fixed once and for all at their physical values. Up- and down-quarks are
light, such that these quantities can be treated as perturbations. Next comes
the strange-quark, whose mass is considerably larger, but nonetheless can be
regarded as a perturbation, as well. The mass of the charmed quark, on the
other hand, is much too large so as to be treated in an analogous manner. The
three lightest quarks, however, may be viewed as perturbations of the
symmetric Lagrangian of massless QCD, which is invariant under chiral
transformations, $\mbox{G = SU(3)}_{R} \times \mbox{SU(3)}_{L}$.

This so-called chiral limit, $m_u, m_d, m_s \to 0$, represents a purely
theoretical abstraction -- chiral symmetry {\it is} explicitly broken in
nature. Likewise, a zero anisotropy field in an antiferromagnet is to be
regarded as an idealized situation, too. As far as the third symmetry breaking
parameter of interest, the external magnetic field, is concerned, the
situation is different, because this quantity represents an {\it external}
field. In a laboratory we can organize a world of our own, for we have the
possibility to tune the strength of ${\vec H}$. In particular, the situation
which is analogous to the fictitious chiral limit or a zero anisotropy field,
can easily be realized: simply switch off the magnetic field. Then, at zero
field strength, the ground state of a ferromagnet exhibits spontaneous
magnetization -- much like an antiferromagnet displays a nonzero staggered
magnetization in the limit ${\vec h}_{A} \to 0$, or massless QCD exhibits
a nonzero quark condensate.

Since the magnetic field can be varied continuously, the effective calculation
is under control: as long as the field strength $|{\vec H}|$ is kept weak,
the effective Lagrangian method is an efficient tool to investigate the low
energy behavior of magnetic systems. Moreover, the fact that the magnetic
field can be tuned, is a major advantage over QCD, where the quark masses are
fixed: it provides us with a new way to accurately determine some of the low
energy constants of the effective theory. Consider, for example, the
magnetization of a ferromagnet and its variation with respect to temperature
and magnetic field. In the effective expansion of this quantity, different low
energy constants will show up. Their values may be unraveled by fitting the
calculated curves to experimental data, which, furthermore, are already
available in condensed matter physics. In particular, at higher orders of the
low energy expansion, where the number of effective coupling constants turns
out to be large, this procedure may be of considerable help.

\section{Summary and Outlook}
\label{sumOut}

The present work deals with the low energy analysis of nonrelativistic systems
which exhibit collective magnetic behavior. The corresponding excitations near
the ground state, the spin waves or magnons, are regarded as Goldstone bosons,
resulting from a spontaneously broken internal symmetry, O(3) $\to$ O(2).
Their properties may be analyzed in the framework of the effective Lagrangian
method, which tackles the phenomenon of spontaneous symmetry breaking from a
unified point of view. The method exploits the symmetry properties of the
underlying theory and formulates the dynamics in terms of Goldstone fields.

At large wavelengths, the microscopic structure of condensed matter systems
does not play a significant role: in the corresponding leading order effective
Lagrangians, the specific properties of the system only manifest themselves in
the numerical values of a few low energy couplings. Rather, our attention is
devoted to the consequences of spontaneous symmetry breaking -- it is the
hidden symmetry which manifests itself at small momenta, dictating the explicit
appearance of the respective low energy phenomenon.

If the ground state of the system fails to be Lorentz-invariant, charge
densities may pick up nonzero vacuum expectation values. In the case of a
ferromagnet, the spontaneous magnetization embodies this possibility, giving
rise to a topological term in the effective Lagrangian, which is not invariant
under the internal symmetry O(3). Ferromagnetic magnons are nonrelativistic
particles, which possess only one polarization state and obey a quadratic
dispersion relation. The low energy theorem concerning spin-wave scattering
indeed displays a structure, which would not be permitted in the relativistic
domain: the corresponding expressions for the scattering amplitude and total
cross section violate Lorentz symmetry. The results obtained are in agreement
with Dyson's pioneering microscopic analysis of a cubic ferromagnet within the
Heisenberg model.

The antiferromagnet, on the other hand, does not exhibit spontaneous
magnetization, such that a topological term is absent in the effective
Lagrangian. In contrast to the low energy excitations in a ferromagnet,
antiferromagnetic magnons are relativistic particles, which follow a linear
dispersion law and possess two polarization states. Much like in the
relativistic domain, the effective Lagrangian is invariant with respect to the
hidden symmetry O(3); moreover, the explicit expression for an antiferromagnet
closely resembles the one referring to massless quantum chromodynamics (QCD).
The T-matrix element, describing antiferromagnetic spin-wave scattering,
unlike for the ferromagnet before, turns out to be Lorentz-invariant -- its
structure is analogous to that of the leading order pion-pion scattering
amplitude in QCD, demonstrating the universal feature of the effective
Lagrangian technique.

In either case, ferro- and antiferromagnetic spin-wave scattering, the
calculation is readily done within the effective framework, to be contrasted
with the microscopic approach, where the corresponding analysis turns out to
be fairly involved. In this respect, the situation is analogous to that of
quantum chromodynamics, where the complicated analysis of pion-pion scattering
by means of current algebra methods has been replaced by the effective
Lagrangian technique. Unlike for the strong interaction, scattering processes
in magnetic systems, unfortunately, are not the suitable tool to experimentally
detect interactions among the Goldstone degrees of freedom.

The present work includes approximate symmetries and discusses the modification
of the low energy structure imposed by explicit symmetry breaking. Two
different perturbations of the isotropic Heisenberg model are considered: a
constant external magnetic and a constant anisotropy field. The former quantity
represents a rather special case, since this symmetry breaking parameter is
coupled to the generators of the group O(3) -- in the effective Lagrangian of a
ferromagnet the magnetic field is associated with the spontaneous magnetization,
while, for the antiferromagnet, it appears in the time component of a covariant
derivative. The anisotropy field, on the other hand, which plays a significant
role in connection with the antiferromagnet, belongs to the more general class
of symmetry breaking parameters which are not coupled to the generators of the
hidden symmetry. It leads to an additional term in the effective Lagrangian,
which is proportional to the staggered magnetization.

The dispersion relations regarding spin waves in the presence of an external
magnetic and an anisotropy field, respectively, are in agreement with the
findings of condensed matter physics. Due to a magnetic field, ferromagnetic
magnons experience an overall shift, while the degeneracy of the two
polarization states of antiferromagnetic magnons is lifted. The anisotropy
field provides antiferromagnetic magnons with a "mass", leading to a formula
analogous to the Gell-Mann--Oakes--Renner relation in QCD.

Remarkably, to the order considered, the scattering process concerning
ferromagnetic spin waves is not affected by a magnetic field. Also, the
T-matrix element, describing the analogous interaction in an antiferromagnet,
does not receive additional terms from the magnetic field. On the other hand,
the anisotropy field modifies the low energy theorem concerning
antiferromagnetic spin-wave scattering, leading to an additional contribution
in the T-matrix element, which is on the same footing as the quark mass term in
the pion-pion scattering amplitude.

The present work demonstrates that the leading order effective Lagrangians
permit a concise and straightforward analysis of the low energy properties of
ferro- and antiferromagnets, above all in connection with spin-wave scattering
processes. The effective machinery may now be transferred to more complicated
applications, such as the investigation of thermodynamic quantities. Indeed,
the low temperature expansion for the partition function of an antiferromagnet
has been calculated to three loops \cite{Hofmann AF}, while the results
concerning the temperature dependence of the spontaneous magnetization of a
ferromagnet will be presented in a forthcoming paper \cite{Hofmann Ferro}.

\acknowledgements
I would like to thank H. Leutwyler for his patient assistance throughout this
work and for his critical reading of the manuscript. Thanks also to G.
Colangelo, P. Hasenfratz, S. Mallik, A. V. Manohar and D. Toublan for their
help. I am greatly indebted to the Janggen-P\"ohn-Stiftung for supporting my
doctoral thesis. Likewise, support by Schweizerischer Nationalfonds is
gratefully acknowledged.


\section*{Appendix A: Spin Waves as Collective Excitations}
\renewcommand{\theequation}{A.\arabic{equation}}
\setcounter{equation}{0}
In this appendix, we develop a semiclassical picture of spin waves, which
regards these low energy excitations as some kind of distortion of the
microscopic spin structure \cite{Nolting zwei,Heller Kramers,Keffer Kaplan
Yafet,Kittel first approach}. Afterwards, we try to illuminate Goldstone's
theorem in the present context. As a first step, we have to construct the
microscopic representation of the one-magnon states.

Instead of the spin operators $S^i_n$, introduced in section \ref{NRD}, we
take the following linear combinations thereof,
\begin{equation}
S^+_n \, = \, S^1_n + iS^2_n , \qquad S^-_n \, = \, S^1_n - iS^2_n ,
\end{equation}
and perform a discrete Fourier transformation:
\begin{equation}
S^{\pm}(\vec k) \, = \, \sum_n exp(i{\vec k}{\vec r}_n) \, S^{\pm}_n .
\end{equation}
Note that the operators $S^{\pm}(\vec k)$ refer to the reciprocal lattice.

Next, we apply these operators to the ground state of a ferromagnet. Since,
in our convention, all of the spins point in the direction of the positive
$3^{rd}$ axis, the spin-raising operator, $S^+(\vec k)$, yields identically
zero, $S^+(\vec k) \,|0 \rangle \equiv 0$. The spin-lowering operator,
$S^-(\vec k)$, however, leads to an eigenstate of the Heisenberg Hamiltonian,
\begin{equation}
| {\vec k} \rangle = \frac{1}{\sqrt{2SN}} \, S^-(\vec k) \,|0 \rangle , \qquad
\langle {\vec k} | {\vec k} \rangle = 1 .
\end{equation}
$S$ is the spin quantum number and $N$ denotes the total number of lattice
sites. Consider now the expectation value of the {\it local} operator $S^3_n$
in this one-magnon state (see e.g. \cite{Nolting zwei}):
\begin{equation}
\langle {\vec k} | \, S^3_n \, | {\vec k} \rangle \ = \ S - 1/N .
\end{equation}
This is quite a remarkable finding, since the right hand side is independent
of the specific lattice site $n$ as well as of the wave vector ${\vec k}$. A
spin deviation of one unit, relative to the ground state, is thus uniformly
distributed over the whole lattice -- per site $n$, the spin deviation from
the totally ordered state equals $1/N$. In this picture, a spin wave
corresponds to some sort of collective excitation moving through the lattice,
characterized by a wave vector ${\vec k}$. Since the total spin deviation
amounts to one unit, these quasiparticles are bosons -- so far so good.

In accordance with the semiclassical vector model, each localized spin which
takes part in such a collective mode precesses around the $3^{rd}$ axis. The 
corresponding opening angle is such that the projection of a particular spin
vector on the $3^{rd}$ axis is given by $S - 1/N$. Moreover, there is a
constant phase difference between any two adjacent spins of the collective
mode, depending on the magnitude of the wave vector ${\vec k}$ -- in
particular, at large wavelengths, $|{\vec k}| \rightarrow 0$, all the spins
precess in phase.

Now, in order to get an intuitive understanding of Goldstone's theorem in the
present context, let us briefly discuss a simple model of a ferromagnet
\cite{Heisenberg}. Consider a linear chain of $N$ spin-$\mbox{$ \frac{1}{2}$}$
vectors, bent around into a ring, such that the first and the ${(N + 1)}^{th}$
spin are identical. Suppose that there is an interaction between adjacent
spins, tending to align them parallel. The ground state, in which all the
spins point in the same direction, is degenerate. From the manifold of these
$N + 1$ lowest lying states let us chose, for definiteness, the one with
maximal spin projection $N/2$ on the $z$-axis. Next, construct a one-magnon
state $| k \rangle$,
\begin{eqnarray}
| k \rangle \, = \, \frac{1}{\sqrt N} \sum_n e^{ikn} \, | n \rangle , \qquad
\ k = \frac{2\pi}{N} \, r , \qquad r = 1, 2, \ldots , \nonumber\\| n \rangle
\, = \, \mbox{$ \frac{1}{2}$} ({\sigma}_{n,x} - i{\sigma}_{n,y}) \,|0 \rangle ,
\hspace{3cm}
\end{eqnarray}
where the quantities ${\sigma}_n$ represent Pauli matrices. Now, for $ k = 0$,
the one-magnon state $| k \rangle$ is energetically degenerate with $\,|0
\rangle$ -- these two states differ, however, with respect to the projection of
their total spin on the third axis: for $| k = 0 \rangle$, one obtains
$N/2 - 1$. In particular, this state may be regarded as another of the possible
ground states, whose "spontaneous magnetization" points in a direction different
from the third axis or, equivalently, with a different value of the third
component of the spin vector along the same direction. These two configurations
are thus related by a symmetry transformation: with a suitable rotation, the
ground state, $\,|0 \rangle$, may be transformed into $| k = 0 \rangle$.

Reasoning by analogy, we may transfer these statements to a three-dimensional
ferromagnet. In particular, such a rotation of the system as a whole would not
require any energy: since the spin structure of the ground state is not
distorted while rotating the rigid spin lattice, there are no restoring forces.
As a consequence, there exists at least one form of elementary excitation, the
one corresponding to ${\vec k} = 0$, which gives a rotation of the entire
system in spin space, and which must have zero frequency, $\omega = 0$
\cite{Anderson solids}. These "(${\vec k} = 0 / \omega = 0$)-excitations"
above the ground state, however, are just other ground states and the real
question rather is, whether there is some sort of excitation with no energy
gap in the {\it limit} ${\vec k} \to 0$ \cite{Lange}. Under one additional
condition, the absence of long-range forces, there will indeed exist a whole
branch in the spectrum of elementary excitations, whose frequency continuously
tends to zero, $\omega \to 0$, in the limit ${\vec k} \to 0$ -- here we
recognize the nonrelativistic version of Goldstone's theorem.

Imagine an external field, causing a slight distortion of the ordered spin
structure, such that the direction of the magnetization slowly varies in
space, periodically over some characteristic length $\lambda$. If the external
field is switched off, the system begins to oscillate with some characteristic
frequency or frequencies. Goldstone's theorem then deals with the question
whether or not such collective modes have an energy gap as the characteristic
length in the original distortion of the order parameter tends to infinity.
Since the summation in the Heisenberg exchange Hamiltonian merely extends over
nearest neighbors, by definition, no long-range forces are present in this
model. Therefore, Goldstone excitations do occur.


\section*{Appendix B: Scattering Amplitude and Cross Section}
\renewcommand{\theequation}{B.\arabic{equation}}
\setcounter{equation}{0}
As is well-known from relativistic quantum mechanics, the cross section
referring to elastic two-particle scattering is given by
\begin{equation}
\label{defTR}
d\sigma \ = \ \frac{{|T^{R}|}^2}{4 \sqrt{(k_1 \! \cdot \! k_2)^2 - m^2_1
\, m^2_2}} \; (2\pi)^4 \, {\delta}^4 (k_1 + k_2 - k_3 - k_4)\,
\frac{d^3k_3}{(2\pi)^3 2{\omega}_3 } \frac{d^3k_4}{(2 \pi)^3 2{\omega}_4} ,
\end{equation}
with
\begin{equation}
{\delta}^4 (k_1 + k_2 - k_3 - k_4) \, = \, \delta ({\omega}_1 + {\omega}_2 -
{\omega}_3 - {\omega}_4) \, {\delta}^3 ({\vec k_1} + {\vec k_2} - {\vec k}_3 -
{\vec k_4}) .
\end{equation}
For collinear collisions, the Lorentz-invariant square root may be expressed by
the relative velocity $|{\vec v}\,|$ of the ingoing particles
\cite{Greiner,Aitchison Hey},
\begin{equation}
{\omega}_1\, {\omega}_2 \, |{\vec v}\,| \, = \, \sqrt{(k_1 \! \cdot \! k_2)^2
- m_1^2 \, m_2^2} , \qquad \quad |{\vec v}\,| \, = \, |{\vec v_1} - {\vec
v_2}| .
\end{equation}
Proceeding this way, we do not specialize to the dispersion law of massive
relativistic particles. Note that a relativistic normalization of the
one-particle states has been used,
\begin{equation}
\langle {\vec k} \, | \, {\vec k'} \rangle \, = \, (2 \pi)^3 2{\omega} \,
{\delta}^3 ({\vec k} - {\vec k'}) \ \ \ \Longleftrightarrow \ \ \ \int \!\!
\frac{d^3k}{(2 \pi)^3 2{\omega}} \, | \, {\vec k} \, \rangle \langle {\vec k}
\, | g
\end{equation}
Likewise, in Born approximation, the quantity $T^{R}$,
\begin{equation}
T^{R} = {\langle {\vec k_3} {\vec k_4} \, | \, {\cal L}_{int} \, | \,
{\vec k_1} {\vec k_2} \rangle}^{R} ,
\end{equation}
is the $T$-matrix element, evaluated in this specific normalization.

Since our analysis of ferro- and antiferromagnets is based on a
nonrelativistic normalization of the one-particle states,
\begin{equation}
\langle {\vec k} \, | \, {\vec k'} \rangle \, = \, (2 \pi)^3 {\delta}^3 ({\vec
k} - {\vec k'}) \ \ \ \Longleftrightarrow \ \ \ \int \!\! \frac{d^3k}{(2
\pi)^3} \, | \, {\vec k} \, \rangle \langle {\vec k} \, | ,
\end{equation}
the expression (\ref{defTR}) has to be modified accordingly: instead of
normalizing to $2{\omega}$ particles in a given volume $V$, we shall
normalize to one particle per volume $V$. Hence, the formula for the
scattering cross section now reads:
\begin{equation}
\label{defTNR}
d\sigma \ = \ \frac{{|T^{NR}|}^2}{|{\vec v}\,|} \; (2 \pi)^4 \,
{\delta}^4 (k_1 + k_2 - k_3 - k_4) \, \frac{d^3k_3}{(2 \pi)^3} \frac{d^3k_4}{(2
\pi)^3} .
\end{equation}
The quantity $T^{NR}$ then represents the $T$-matrix element
\begin{equation}
T^{NR} = {\langle {\vec k_3} {\vec k_4} \, | \, {\cal L}_{int} \, | \,
{\vec k_1} {\vec k_2} \rangle}^{NR} ,
\end{equation}
where the one-particle states as well as the field operators, appearing in the
interaction Lagrangian, are normalized nonrelativistically.

We are now going to derive the explicit expression for the differential cross
section, assuming that the interacting particles obey a quadratic dispersion
relation, $\omega = \gamma {\vec k}^2$. In connection with condensed matter,
it is appropriate to consider the specific frame of reference, which is given
by the solid body at rest. Accordingly, the initial particle configuration is
to be characterized by two arbitrary wave vectors, ${\vec k}_1$ and ${\vec
k}_2$. It is convenient to choose ingoing and outgoing momenta as follows,
\begin{eqnarray}
{\vec k}_1 & = & \mbox{$ \frac{1}{2}$} ({\vec K} + {\vec q}\,) , \qquad \;
{\vec k}_2 = \mbox{$ \frac{1}{2}$} ({\vec K} - {\vec q}\,) , \nonumber\\
{\vec k}_3 & = & \mbox{$ \frac{1}{2}$} ({\vec K} + {\vec q}\,') ,
\qquad {\vec k}_4 = \mbox{$ \frac{1}{2}$} ({\vec K} - {\vec q}\,') , 
\end{eqnarray}
where the vectors ${\vec K}$ and ${\vec q}$,
\begin{equation}
{\vec K} = {\vec k}_1 + {\vec k}_2 , \qquad {\vec q} = {\vec k}_1 - {\vec
k}_2 ,
\end{equation}
represent total and relative momentum, respectively. In the case of quadratic
kinematics, conservation of energy and momentum leads to $|{\vec q}\,| = |{\vec
q}\,'|$. The scattering angle $\vartheta$ is chosen as the angle
between the direction of ${\vec q}\,'$ relative to that of ${\vec q}$. In
these coordinates, the evaluation of the phase space integral is readily done,
resulting in
\begin{equation}
\int \! \! {\delta}^4 (k_1 + k_2 - k_3 - k_4) \, d^3k_3 \, d^3k_4 \, \to \,
\frac{1}{8\gamma} \, |{\vec q}\,| \, d\Omega .
\end{equation} 
With the relative velocity of the ingoing particles,
\begin{equation}
|{\vec v}\,| \, = \, 2 \gamma \, |{\vec k}_1 - {\vec k}_2| \, = \, 2 \gamma
|{\vec q}\,| ,
\end{equation}
the expression for the differential cross section then amounts to
\begin{equation}
{\frac{d\sigma}{d\Omega}} \ = \ \frac{1}{128 {\pi}^2 {\gamma}^2} \;
{|T^{NR}|}^2 .
\end{equation}
Note that a factor of one half has been included in the above formula,
considering the fact that the interacting particles are identical.


\begin{references}

\bibitem[*]{Adr}
On leave from Institute for Theoretical Physics, University of Bern,
Sidlerstrasse 5, CH-3012 Bern, Switzerland.

\bibitem{Weinberg junior}
S. Weinberg, Phys. Rev. Lett. {\bf{18}}, 188, 507 (1967); Phys. Rev.
{\bf 166}, 1568 (1968);\\
R. Dashen, Phys. Rev. {\bf{183}}, 1245 (1969);\\
R. Dashen and M. Weinstein, Phys. Rev. {\bf{183}}, 1261 (1969).

\bibitem{Coleman Callan}
S. Coleman, J. Wess and B. Zumino, Phys. Rev. {\bf{177}}, 2239 (1969);\\
C. G. Callan, S. Coleman, J. Wess and B. Zumino, Phys. Rev. {\bf{177}}, 2247
(1969).

\bibitem{Li Pagels}
L.-F. Li and H. Pagels, Phys. Rev. Lett. {\bf 26}, 1204 (1971);\\
H. Pagels, Phys. Rep. C {\bf 16}, 219 (1975).

\bibitem{Weinberg seminal}
S. Weinberg, Physica A {\bf 96}, 327 (1979).

\bibitem{Gasser Leutwyler QCD}
J. Gasser and H. Leutwyler, Ann. Phys. (N.Y.) {\bf{158}}, 142 (1984);
Nucl. Phys. B {\bf 250}, 465 (1985).

\bibitem{Hasenfratz Leutwyler}
P. Hasenfratz and H. Leutwyler, Nucl. Phys. B {\bf 343}, 241 (1990).

\bibitem{Hasenfratz Niedermayer 1991}
P. Hasenfratz and F. Niedermayer, Phys. Lett. B {\bf 268}, 231 (1991).

\bibitem{Hasenfratz Niedermayer 1993}
P. Hasenfratz and F. Niedermayer, Z. Phys. B {\bf 92}, 91 (1993).

\bibitem{Leutwyler NRD}
H. Leutwyler, Phys. Rev. D {\bf 49}, 3033 (1994).

\bibitem{Leutwyler Phonons}
H. Leutwyler, Helv. Phys. Acta {\bf 70}, 275 (1997).

\bibitem{Soto}
R. M. Roman and J. Soto, {\it Effective Field Theory Approach to Ferromagnets
and Antiferromagnets in Crystalline Solids}, preprint Univ. of Barcelona
UB-ECM-PF 97/23, cond-mat/9709298.

\bibitem{Burgess}
C. P. Burgess, {\it Goldstone and Pseudogoldstone Bosons in Nuclear, Particle
and Condensed Matter Physics}, hep-th/9808176.

\bibitem{Leutwyler foundations}
H. Leutwyler, Ann. Phys. {\bf{235}}, 165 (1994).

\bibitem{Fradkin}
E. Fradkin, {\it Field Theories of Condensed Matter Systems}, Frontiers in
Physics Vol. 82 (Addison-Wesley, 1991).

\bibitem{Dyson1}
F. J. Dyson, Phys. Rev. {\bf{102}}, 1217 (1956).

\bibitem{Leutwyler Brasil}
H. Leutwyler, in {\it Hadron Physics 94 -- Topics on the Structure and
Interaction of Hadronic Systems}, edited by V. E. Herscovitz, C. A. Z.
Vasconcellos and E. Ferreira (World Scientific, 1995), p. 1.

\bibitem{footnote1}
A recent analysis, considering the manifestation of the crystal point group
${\bar 3}m$ at next-to-leading order in the effective Lagrangian for ferro-
and antiferromagnets, can be found in \cite{Soto}.

\bibitem{footnote2}
We identify the zeroth component of the coordinate vector with time: $x^0 = t$
(no factor of c).

\bibitem{Goldstone}
J. Goldstone, Nuovo Cim. {\bf 19}, 154 (1961);\\
J. Goldstone, A. Salam and S. Weinberg, Phys. Rev. {\bf 127}, 965 (1962).

\bibitem{Coleman}
S. Coleman, Erice Lectures 1973, in {\it Laws of Hadronic Matter} (Academic
Press, London and New York, 1975); reprinted in S. Coleman, {\it Aspects of
Symmetry} (Cambridge Univ. Press, 1985).

\bibitem{Guralnik Hagen Kibble}
G. S. Guralnik, C. R. Hagen and T. W. B. Kibble, in {\it Advances in Particle
Physics}, edited by R. L. Cool and R. E. Marshak (Wiley, New York, 1968), Vol.
2, p. 567.

\bibitem{footnote3}
The phase of the Goldstone-boson states $|{\pi}^a(\vec k)\rangle$ may be chosen
such that the constant $F$ is real and positive.

\bibitem{Lange}
R. V. Lange, Phys. Rev. Lett. {\bf 14}, 3 (1965); Phys. Rev. {\bf 146}, 301
(1966).

\bibitem{Chadha Nielsen}
H. B. Nielsen and S. Chadha, Nucl. Phys. B {\bf 105}, 445 (1976).

\bibitem{footnote4}
The zeroth component of the momentum vector is identified with the energy,
$k^0 = \omega$ (no factor of c); moreover, $kx \equiv \omega t -
{\vec k}{\vec x}$.

\bibitem{Kittel QT}
C. Kittel, {\it Quantum Theory of Solids} (Wiley, New York, 1987).

\bibitem{Landau Lifshitz}
L. D. Landau and E. M. Lifshitz, {\it Statistical Physics}, edited by E. M.
Lifshitz and L. P. Pitajewski, Course of Theoretical Physics Vol. 9 (Pergamon,
London, 1981), Part 2.

\bibitem{Randjbar-Daemi Salam Strathdee}
S. Randjbar-Daemi, A. Salam and J. Strathdee, Phys. Rev. B {\bf 48}, 3190
(1993).

\bibitem{D'Hoker Weinberg}
E. D'Hoker and S. Weinberg, Phys. Rev. D {\bf 50}, 6050 (1994).

\bibitem{footnote5}
For a more detailed account and, particularly, for a derivation of the results
given here, the reader may consult reference \cite{Leutwyler NRD}.

\bibitem{Colangelo et al}
J. Bijnens, G. Colangelo, G. Ecker, J. Gasser and M. E. Sainio, Phys. Lett. B
{\bf 374}, 210 (1996).

\bibitem{footnote6}
More precisely, QCD with an exact chiral symmetry, $\mbox{G = SU(2)}_{R}
\times \mbox{SU(2)}_{L}$ broken down to $\mbox{H = SU(2)}_{V}$.

\bibitem{footnote7}
Since the effective representation of the current (\ref{Jreff})
is proportional to ${\varepsilon}_{ijk}$, and the corresponding transition
matrix element (\ref{AFmatrix}) involves a Kronecker delta, the tensor
${\varepsilon}_{ab}$ will appear either in the definition of the field
operators or the one-magnon-states. We choose the latter possibility.

\bibitem{Keffer Loudon}
F. Keffer and R. Loudon, J. Appl. Phys. (Suppl.) {\bf 32}, 2 (1961).

\bibitem{Oguchi}
T. Oguchi, Phys. Rev. {\bf 117}, 117 (1960).

\bibitem{Beeman Pincus}
D. Beeman and P. Pincus, Phys. Rev. {\bf 166}, 359 (1968).

\bibitem{Brooks Harris}
A. Brooks Harris, Phys. Rev. {\bf 183}, 486 (1969).

\bibitem{Kashcheev}
V. N. Kashcheev, Soviet Phys. -- Solid State {\bf 4}, 556 (1962).

\bibitem{footnote8}
Note that a {\it relativistic} normalization
of the one-pion states has been used, $\langle {\pi}^a (\vec k) \, | \,
{\pi}^b (\vec k') \rangle = (2{\pi})^3 \, {\delta}^{ab} \, 2k^0 \,
{\delta}^3({\vec k} - {\vec k'})$. $k^0$ is the time component of the momentum
four vector, $k^{\mu} = ({\omega}/c, {\vec k})$.

\bibitem{footnote9}
The whole analysis still refers to leading order of the effective expansion --
the low energy theorems to be derived below receive corrections due to higher
order terms, i.e. terms of higher orders in the momenta than ${\vec k}^2$.

\bibitem{footnote10}
In order to avoid confusion with signs, the magnetic moment $\mu$ is taken
positive: the spin vectors and the spontaneous magnetization, respectively,
thus point along the {\it same} direction as the external magnetic field. This
convention will be maintained throughout the paper.

\bibitem{Nolting zwei}
W. Nolting, {\it Quantentheorie des Magnetismus} (Teubner, Stuttgart,
1986), Band 2.

\bibitem{van Kranendonk van Vleck}
J. van Kranendonk and J. H. van Vleck, Rev. Mod. Phys. {\bf 30}, 1 (1958).

\bibitem{Keffer}
F. Keffer, {\it Spin Waves}, in {\it Encyclopedia of Physics --
Ferromagnetism}, edited by S. Fl\"ugge and H. P. J. Wijn (Springer, Berlin,
1966), Vol. 18-2, p. 1.

\bibitem{Tjablikow}
S. W. Tjablikow, {\it Quantentheoretische Methoden des Magnetismus} (Teubner,
Leipzig, 1968).

\bibitem{Herpin}
A. Herpin, {\it Th\'eorie du Magn\'etisme} (Presses Universitaires de France,
Paris, 1968).

\bibitem{footnote11}
Note that, for antiferromagnets, the magnetic field ${\vec
H}$ occurs on the same footing as the wave vector ${\vec k}$, $f^i_0(x)
\propto |{\vec k}|$, whereas the anisotropy field ${\vec h}_{A}$ counts as
quantity of order two, $m_{\alpha}(x) \propto {\vec k}^2$.

\bibitem{Gell-Mann Oakes Renner}
M. Gell-Mann, R. J. Oakes and B. Renner, Phys. Rev. {\bf 175}, 2195 (1968).

\bibitem{Wagner}
D. Wagner, {\it Introduction to the Theory of Magnetism} (Pergamon, Oxford,
1972).

\bibitem{Hofmann AF}
C. P. Hofmann, {\it Effective Analysis of the O(N) Antiferromagnet: Low
Temperature Expansion of the Order Parameter}, preprint Univ. Bern
BUTP-97/15, hep-ph/9706418, submitted to Phys. Rev. B.

\bibitem{Hofmann Ferro}
C. P. Hofmann, in preparation.

\bibitem{Heller Kramers}
G. Heller and H. A. Kramers, Proc. Roy. Acad. Amsterdam {\bf 37}, 378 (1934).

\bibitem{Keffer Kaplan Yafet}
F. Keffer, H. Kaplan and Y. Yafet, Amer. J. Phys. {\bf 21}, 250 (1953).

\bibitem{Kittel first approach}
C. Kittel, {\it Introduction to Solid State Physics} (Wiley, New York, 1986).

\bibitem{Heisenberg}
W. Heisenberg, {\it Einf\"uhrung in die einheitliche Feldtheorie
der Elementarteilchen} (Hirzel, Stuttgart, 1967).

\bibitem{Anderson solids}
P. W. Anderson, {\it Concepts in Solids} (Benjamin, New York, 1963).

\bibitem{Greiner}
W. Greiner and J. Reinhardt, {\it Quantenelektrodynamik}, Theoretische Physik
Band 7 (Harri Deutsch, Thun, 1984).

\bibitem{Aitchison Hey}
I. J. R. Aitchison and A. J. G. Hey, {\it Gauge Theories in Particle Physics}
(A. Hilger, Bristol, 1982).


\end{references}
\end{document}